\documentclass{jfm}

\usepackage{graphicx,amsmath,stmaryrd}
\usepackage{natbib}
\usepackage{color}

\usepackage{bm,textcomp}
\usepackage{ulem}

\ifCUPmtlplainloaded \else
  \checkfont{eurm10}
  \iffontfound
    \IfFileExists{upmath.sty}
      {\typeout{^^JFound AMS Euler Roman fonts on the system,
                   using the 'upmath' package.^^J}%
       \usepackage{upmath}}
      {\typeout{^^JFound AMS Euler Roman fonts on the system, but you
                   dont seem to have the}%
       \typeout{'upmath' package installed. JFM.cls can take advantage
                 of these fonts,^^Jif you use 'upmath' package.^^J}%
      }
  \else
  \fi
\fi

\ifCUPmtlplainloaded \else
  \checkfont{msam10}
  \iffontfound
    \IfFileExists{amssymb.sty}
      {\typeout{^^JFound AMS Symbol fonts on the system, using the
                'amssymb' package.^^J}%
       \usepackage{amssymb}%
       \let\le=\leqslant  
         
      }{}
  \fi
\fi

\ifCUPmtlplainloaded \else
  \IfFileExists{amsbsy.sty}
    {\typeout{^^JFound the 'amsbsy' package on the system, using it.^^J}%
     \usepackage{amsbsy}}
    {\providecommand\boldsymbol[1]{\mbox{\boldmath $##1$}}}
\fi

\newsavebox{\astrutbox}
\sbox{\astrutbox}{\rule[-5pt]{0pt}{20pt}}

\title[Red blood cells in  shear flow]{Dynamics of a large population of red blood cells under shear flow}

\author[C. Minetti , V. Audemar, T. Podgorski and G. Coupier]
{C.\ns M\ls I\ls N\ls E\ls T\ls T\ls I$^1$,\ns  V.\ns A\ls U\ls D\ls E\ls M\ls A\ls R$^{2}$,\ns T.\ns P\ls O\ls D\ls G\ls O\ls R\ls S\ls K\ls I$^{2}$ \and G.\ns C\ls O\ls U\ls P\ls I\ls E\ls R$^{2}$}

\affiliation{$^1$ Service de chimie physique EP, Universit\'e libre de Bruxelles, 50, avenue Frankin-Roosevelt, CP16/62, B-1050 Brussels, Belgium. $^2$ Universit\'e Grenoble Alpes, CNRS, LIPhy, F-38000 Grenoble, France.}

\pubyear{}
\volume{}
\pagerange{}
\date{?? and in revised form ??}

\begin{document}

\maketitle 

\begin{abstract}

An exhaustive description of the dynamics  under shear flow  of a large number of red blood cells in dilute regime is proposed, which highlights and takes into account the dispersion in cell properties within a given blood sample. Physiological suspending fluid viscosity is considered, a configuration surprisingly seldom considered in experimental studies, as well as a more viscous fluid that is a reference in the literature. Stable and unstable flipping motions well described by Jeffery orbits  or modified Jeffery orbits are identified, as well as transitions to and from tank-treading motion in the more viscous suspending fluid case. Hysteresis loops upon shear rate increase or decrease are highlighted for the transitions between unstable and stable orbits as well as for the transition between flipping and tank-treading. We identify which of the characteristic parameters of motion and of the transition thresholds depend on flow stress only or also on suspending fluid viscosity.

\end{abstract}


\section{Introduction}

In human blood, the haematocrit, that is, the volume fraction of red blood cells (RBCs) is usually between
40 and 50\% while other cells, platelets and  white blood cells, occupy
less than 1\% of blood volume. Under healthy conditions, RBCs are highly
deformable cells. This allows them to easily squeeze through narrow
capillaries while preventing clogging even at high volume fraction, and
to migrate away from vessel walls, which optimizes transport thanks to the
formation of a lubricating cell-free layer. Cells also contribute strongly to blood viscosity 
and its shear-thinning behaviour
 \cite[see][]{chien70,pries92,vitkova08,forsyth11,fedosov11}, which allows to partly 
compensate for the higher heart power
needed in situations of effort. While in most situations blood flow is
influenced by interactions with walls and between cells, the question of
the dynamics of a single RBC in a simple linear shear flow is of utmost
fundamental interest and has been the subject of many experimental and
theoretical studies due to its complexity and its sensitivity to RBC
properties. From a structural and mechanical viewpoint, a RBC is
fundamentally a liquid drop encapsulated in a membrane. The cytosol is a
haemoglobin solution with an average concentration (MCHC: Mean
Corpuscular Haemoglobin Concentration) normally between 320 and 360 g/l 
corresponding to a mean viscosity in the range 6-10 mPa.s \cite[see][]{ross77}. 
The RBC membrane is composed of a lipid
bilayer that provides area dilation resistance and bending rigidity,
and a spectrin network that constitutes a quasi 2D skeleton on the
inner surface of the lipid bilayer and connected to it via junction
complexes and membrane proteins. This skeleton provides 2D shear
elasticity to the membrane and to some extent shape memory. The cell
volume is about 90 $\mu$m$^3$ on average and the membrane area 136 $\mu$m$^2$,
corresponding to a reduced volume of around 0.6  \cite[see][]{linderkamp83,fung93}. 
This leads to the characteristic biconcave disk shape and provides enough excess area
(compared with a sphere) to allow large deformations.

Simple shear flow is a basic rheometric configuration and as such the 
dynamics of a suspended object can be considered as a marker of rheological 
and mechanical properties. In addition, it is also a benchmark situation 
for comparing experiments and theoretical or numerical modelling of the 
red blood cell. Indeed, there are still many open questions about the 
mechanical structure and properties of the RBC: what are the respective roles
of the bending rigidity of the lipid bilayer, the elasticity of the 
spectrin network and the viscosity of the cytosol? Does the cytoskeleton 
have a permanent stress-free shape or should the possibility of remodelling
be considered (e.g. due to NO \cite[see][]{simmonds13,grau13} or ATP  \cite[see][]{betz09} 
release and production in response to stress)? What is this stress-free shape 
\cite[see][]{fischer81,svelc12,peng14,cordasco14,sinha15}? RBC mechanical properties 
are an indicator and consequence of several pathologies (sickle cell disease, thalassaemia, 
elliptocytosisÉ) and can be modified by physical activity or conditions such as long-term 
space flight for instance \cite[see][]{rizzo12}. 
It has been shown that these conditions can induce modifications of membrane composition
and properties, RBC shape and internal viscosity. The corresponding variations of dynamics, 
deformation and orientation of RBCs in flow strongly condition
blood rheology (viscosity, viscoelasticity) and the hydrodynamic
interactions that govern the structure of blood flows in vessels through lift forces
near walls, interactions between cells and distribution in networks \cite[see][]{grandchamp13,shen16,roman16}.

Numerous studies, both experimental \cite[see][]{morris79,goldsmith72,bitbol86,abkarian07,
dupire12,fischer13,levant16,lanotte16,mauer18} and numerical \cite[see][]{cordasco13,cordasco14,
peng14,sinha15,lanotte16,mauer18} have been devoted to the dynamics of a single RBC in a 
shear flow. Efforts to capture the main features of the dynamics are recent \cite[see][]{dupire15,mendez18}.  In an attempt to isolate the contributions of shear elasticity
and membrane bending energy, investigations related to the modelling of blood flows
using elastic capsules  or giant lipid vesicles
as simplified models of RBCs already show rather complex diagrams of dynamical 
states notably involving tumbling (TB), tank-treating (TT), vacillating-breathing (VB) 
 modes as a function of membrane properties, viscosities of the external and internal 
media, shear rate and rest shape of the object \cite[see][]{barthes-biesel81,ramanujan98,lac05,skotheim07,kessler09,bagchi09,walter11,foessel11,dupont13,dupont16,barthes-biesel16,dehaas97_1,rioual04,kantsler05,abkarian05,noguchi05_3,noguchi05_1,kantsler06,mader06,misbah06,mader07,noguchi07,lebedev07,Danker07b,kantsler08,deschamps09b,farutin10,biben11,zabusky11,farutin12_1,farutin12_2,laadhari12}. These parameters are usually combined
in a set of dimensionless numbers such as reduced volume $\nu$, viscosity ratio $\lambda$ 
and capillary number $Ca$, the latter  comparing the hydrodynamic shear with either
bending rigidity or shear elasticity of the membrane. In high viscosity media, the 
drop-like tank-treading motion and its characteristics (cell elongation, inclination 
and tank-treading frequency) have often been put forward as a means to characterize 
RBC mechanical properties \cite[see][]{fischer78,fischer07,chien87}. While other regimes 
that exist for physiological values of viscosity and shear rate, such as 
flipping/tumbling \cite[see][]{goldsmith72}, rolling motion \cite[see][]{bitbol86} or swinging 
motion at intermediate stress and viscosity \cite[see][]{abkarian07} have been identified, 
as well as hysteretic behaviour in the transitions between different 
modes \cite[see][]{dupire12}, there is still no clear consensus on the whole dynamical 
diagram of RBCs in shear flow as a function of shear stress and external viscosity.

Part of the discrepancies and uncertainties on the boundaries of the different
dynamical modes in phase space in experimental studies available 
in the literature is likely to be due to the variability of RBC properties between 
different subjects and even within a given blood sample. Indeed, most studies focused
on the analysis of single cells, sometimes the same cell (by varying shear rate) and 
often different cells when it is necessary to vary the viscosity of suspending medium.
However, RBCs are filled with a haemoglobin solution whose concentration - and
viscosity - can vary significantly during the RBC lifespan, as they tend
to slightly dehydrate when getting older, and between individuals. For instance, the
volume of RBCs can vary by 25 \% between young and old cells \cite[see][]{linderkamp82}, leading 
to equivalent variations of the haemoglobin concentration \emph{at the cell level} (32 to 40 g/dl). 
This leads to dispersion of internal viscosities within a blood sample between 6 and 20 mPa.s 
at 37$^\circ$C \cite[see][]{ross77} and dispersion of mechanical properties and 
dynamics \cite[see][]{pfafferott85}.

In this paper, after summarizing the data on transition thresholds between different 
dynamical modes available in the literature, we report on measurements of the distribution 
of orientation angles and 
aspect ratios of RBCs in large samples by varying suspending medium viscosity and shear 
rate. This allows to derive and quantify the populations of cells that are in the different dynamic modes
for a given set of parameters, for a given healthy blood sample in which natural variability
of cell properties is present. We give ranges of values for the transition thresholds and 
their hysteretic behaviour, that take this variability into account. The large populations 
analyzed for each point of the parameter space (several hundreds of RBCs) provide for the 
first time relevant statistical information on the dispersity of RBC dynamics in shear flow.

We focus on two suspending media: one has high viscosity (25 mPa.s), and allows to make 
comparisons with the literature, which has notably focused on such a configuration because it allows  the tank-treading regime to be reached. The other has a viscosity 
1.5 mPa.s close to that of plasma and has an obvious interest for physiological issues. This situation has been much less studied in the literature, as far as experiments are concerned.

\section{Cell orientation and Jeffery orbits}

We introduce the notations for angles that will be used all along the paper, and Jeffery orbits, that describe the motion of rigid ellipsoids under shear flow. This motion that will be used here in our modelling is also a reference case.

\subsection{Euler angles}
\label{sec:euler}

\begin{figure}
\begin{center}
  \includegraphics[width=0.7\columnwidth]{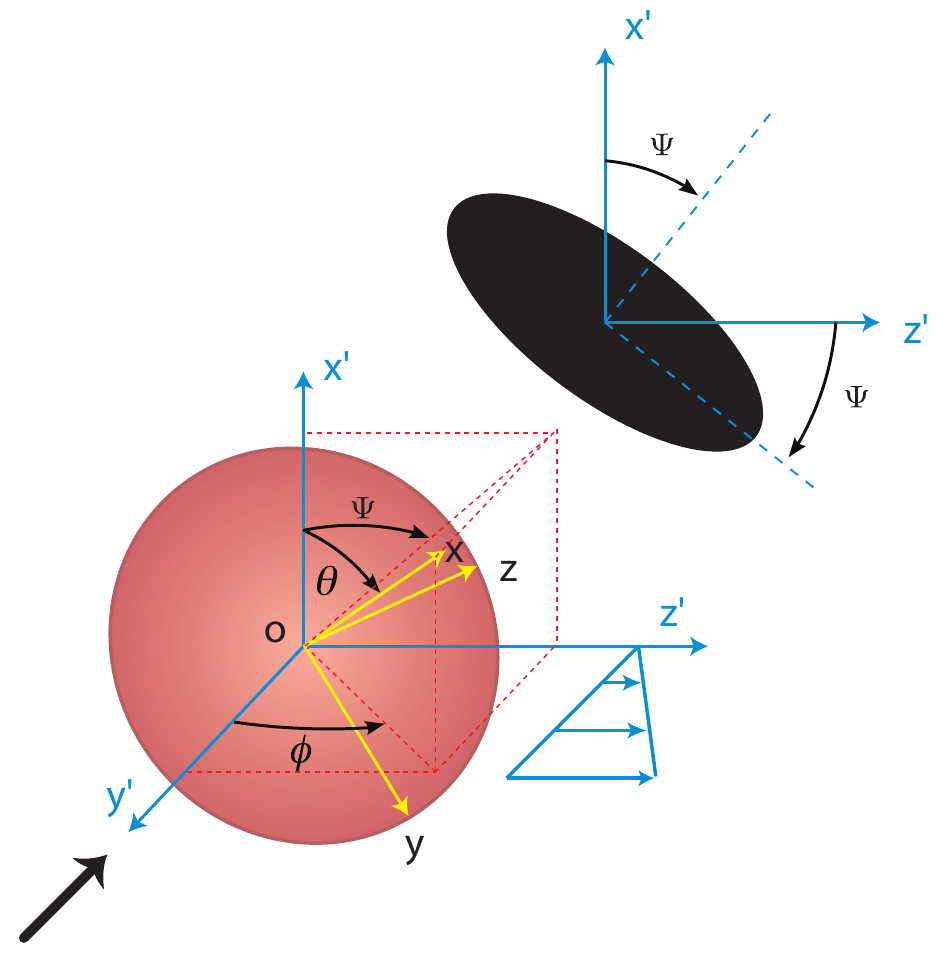}
\caption{The Euler angles used to describe the flipping regime. Convention used  in \cite{jeffery22} and in this paper. The large black arrow shows the viewing direction in the experiments, where projection on the $x'z'$ plane is seen (black ellipse).\label{fig:schema}}
\end{center}
\end{figure}

We consider cells placed in a shear flow with flow direction $Oz'$, shear gradient direction $Oy'$ and vorticity axis $Ox'$. To characterize the orientation and motion of cells, we shall use the Euler angles. Let us consider, as a first approximation of RBC shape, an ellipsoid of equation \begin{equation}\label{eqellips} r^2 x^2+y^2+z^2=1.\end{equation} $r$ is the aspect ratio, which will be larger than 1 here (oblate ellipsoid). This ellipsoid may rotate and we use the Euler angles as defined in \cite{jeffery22} and in Fig. \ref{fig:schema} to describe this rotation in the fixed coordinate system $Ox'y'z'$ that coincides initially with the system $Oxyz$ associated with the ellipsoid (see supplemental material for a comparison with the other convention used in the literature).

In the original paper by Jeffery (\cite{jeffery22}), $\theta$ is obtained by rotation around the $Oz'=Oz$ axis, then $\phi$ is obtained by rotation around the $Ox'$ axis, such that it is defined as the angle between the planes $Ox'y'$ and $Ox'x$ (see Fig. \ref{fig:schema}). With this convention, when $\phi=0$ and $\theta=90^\circ$, the cell face is in the $Ox'z'$ plane.

If the cells are viewed from the velocity gradient axis $Oy'$, as in our experiments, the angle $\Psi$ defined as the angle between $Ox'$ and the  projection of the cell axis of revolution $Ox$ onto the plane $Ox'z'$, can be easily determined (see. Fig. \ref{fig:schema}). This angle is also the angle between the projection of the cell and the flow direction. $\Psi$ is related to $\theta$ and $\phi$ through $\tan \Psi = \tan \theta  \sin \phi$.

\begin{figure}
\begin{center}
  \includegraphics[width=\columnwidth]{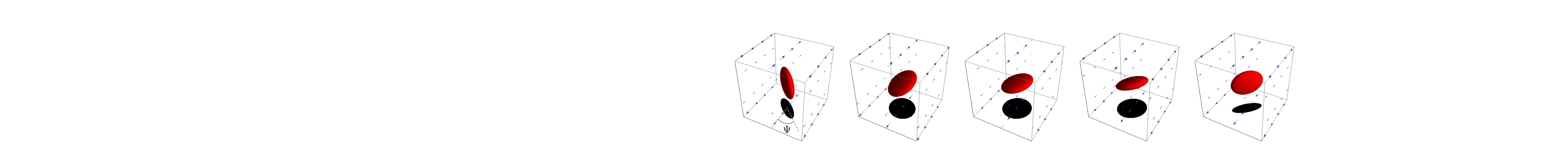}
\caption{Selected snapshots along half a flipping period of an ellipsoid of aspect ratio $r=3$ following a Jeffery orbit of orbit angle $\theta_0=45^\circ$. The black shape is the projection of the ellipsoid parallel to the shear gradient direction; it corresponds to what is seen in our experiment. $\Psi$ is the angle of the main axis of this projected shape with the flow direction. As stated by Eq. \ref{eq:PsiJeff}, it oscillates between $-\theta_0$ and $\theta_0$. The apparent aspect ratio $r_a$ oscillates between a minimal value strictly larger than 1 and $r$, when the cell is seen edge on and $\Psi=\pm\theta_0$.\label{fig:exJeff}}
\end{center}
\end{figure}

\subsection{Jeffery orbits}

For a shear flow in the $z'$ direction with $y'$ the shear gradient direction and $x'$ the vorticity direction, the motion of a rigid  ellipsoid in creeping flow is given by (\cite{jeffery22}):
\begin{eqnarray}
\dot{\theta}&=&\dot{\gamma} \frac{r^2-1}{r^2+1} \sin \theta \cos\theta \sin\phi\cos\phi,\\
\dot{\phi}&=& \frac{\dot{\gamma}}{1+r^{-2}}(r^{-2}\cos^2 \phi+ \sin^2 \phi), \label{eq:vitphi}
\end{eqnarray}

with $\mathrm{C}\equiv r \tan\theta_0$ the orbit parameter. These equations can be solved and give:

\begin{eqnarray}
\tan \theta &=&\frac{\mathrm{C}}{r(r^{-2}\cos^2\phi+\sin^2\phi)^{1/2}},\\
\tan \phi&=&  r^{-1} \,\tan \frac{\dot{\gamma} t}{r+r^{-1}}.\label{eq:phi}
\end{eqnarray}

$\theta$ oscillates between $\theta _0$ and $\arctan C$ (spinning motion).

We shall refer to these possible motions as flipping motions. Among them, $\theta _0=0^\circ$ corresponds to what is called rolling motion  ($\theta$ always equal to 0), while $\theta _0=90^\circ$ corresponds  to tumbling ($\theta$ always equal to $90^\circ$).  Note that when $r>1$, $\dot{\phi}$ is minimal when $\phi=0$, which corresponds to the cell aligned with the flow direction.

Simple trigonometry then yields the equation for $\Psi(t)$ according to Jeffery theory: \begin{equation}\label{eq:PsiJeff} \tan \Psi = \tan \theta_0 \times  \sin \frac{\dot{\gamma} t}{r+r^{-1}}.\end{equation}

$\Psi$ therefore oscillates between two extreme positions $-\theta_0$ and $\theta_0$, where it stays more time (see Fig. \ref{fig:exJeff} for an example).\\

In this paper, we shall consider the possible flipping motions for a cell, that will be hypothesized to closely follow Jeffery orbits, but also the tank-treading motion. In that case, the cell small axis remains in the shear plane  ($\theta=90^\circ$) while the cell adopts a constant angle relatively to the flow direction. This definition is unambiguous as long as the cell does not deform. If it does, as pointed out in \cite{dupont16} where oblate capsules are considered, one should refer to the membrane material point  that was located, at rest, on the small axis. From this unambiguous definition, a cell of fixed shape in flow for which this point is directed toward the vorticity direction would be called a rolling cell, even though its deformation is such that  it ressembles a tank-treading cell. From this point of view, the tank-treading motion mentioned in \cite{cordasco14_2}, Fig.12 or in \cite{sinha15}, Fig. 15 should be called rolling motion. 

Since the numerical studies we will refer to in the following have used the term tank-treading, and since in most experimental studies the deformation seems to remain weak, we shall however go on using the (potentially improper) term of tank-treading for cells whose small axis of symmetry lies in the shear plane. 
 
\section{State of the art}

In the whole paper and in this section in particular, we shall use the Jeffery notations and conventions (Fig. \ref{fig:schema}). The original notations used in the considered papers (which are quite varied)  are recalled in the supplemental material, for sake of clarity. Experiments were run at laboratory temperature, at which the viscosity of the hemoglobin solution is close to 10 mPa.s, when it is equal to 6.5 mPa.s at body temperature. In experimental papers, the "natural" parameter space $(\eta_0,\dot{\gamma})$ is often used, where $\eta_0$ is the carrying fluid viscosity and $\dot{\gamma}$ the shear rate, a parameter that can be varied continuously. As an alternative to $\dot{\gamma}$, the typical  stress on the cell $\tau=\eta_0\dot{\gamma}$ is also often considered. We shall stick to this latter choice in the following.

\begin{figure}
\begin{center}
  \includegraphics[width=\columnwidth]{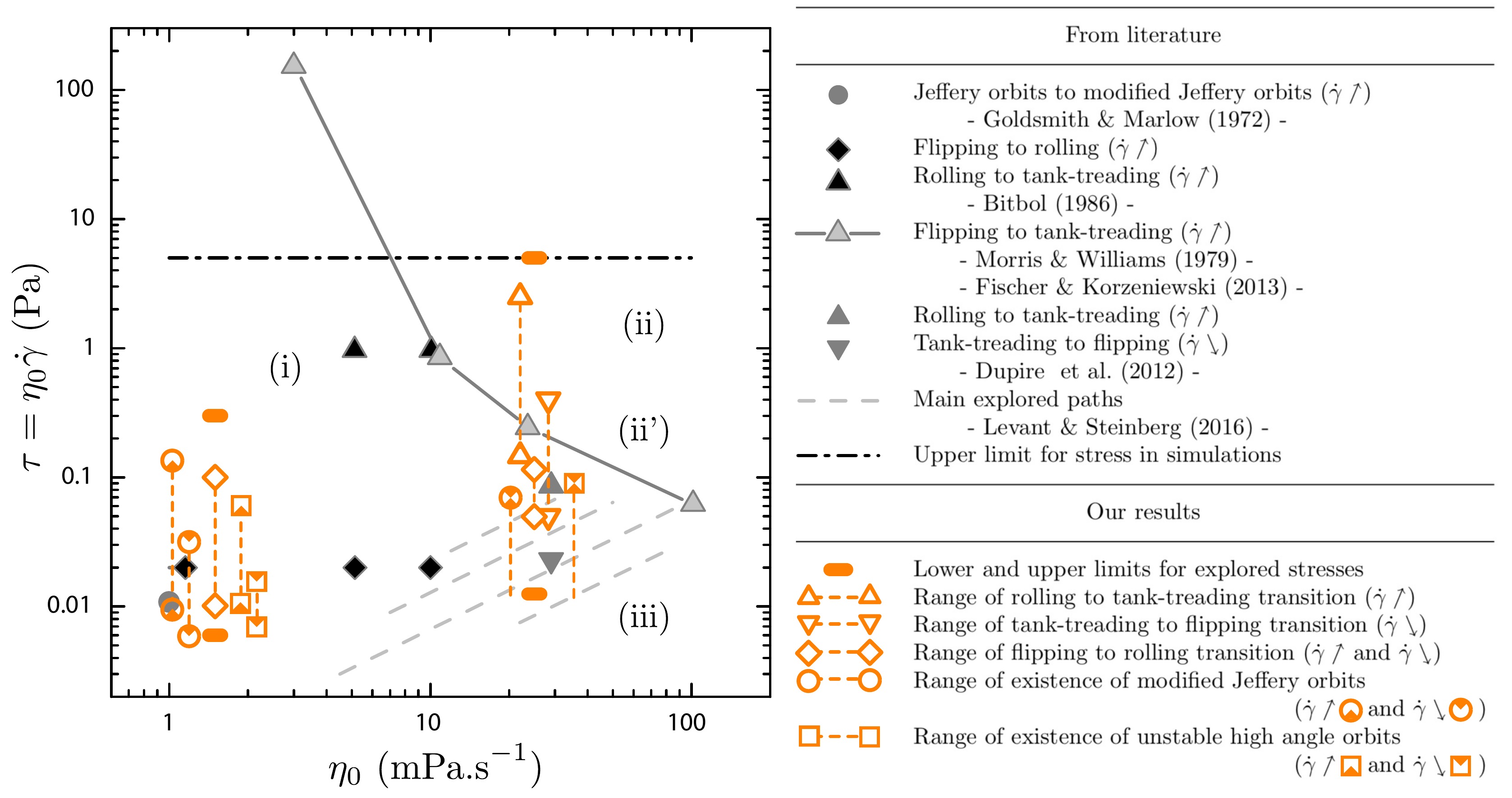}

\caption{Summary of the transition thresholds identified in the literature and in our experiments. Whether the mentioned transition should be understood to occur in the increasing or decreasing shear rate case is indicated by the symbols ($\dot{\gamma} \nnearrow$) and ($\dot{\gamma} \ssearrow$), respectively. The chosen data points from the literature are commented in the text. Due to variability between cells, all of our thresholds are identified by an interval.  Some transitions have no lower bond because of the range of stresses explored. The transition intervals are slightly shifted around the two explored viscosities $1.5$ and 25 mPa.s for clarity. \label{fig:basicdiagram}}
\end{center}
\end{figure}

The main thresholds and dynamics states domains  found in previous experiments from the literature are reported in Fig. \ref{fig:basicdiagram}.

In numerical simulations, dimensionless parameters are naturally chosen. These are $\lambda$ and $C_a$. $\lambda$ is the viscosity contrast that is, the ratio between the viscosity of the haemoglobin solution and that of the carrying fluid. $\lambda=1$ therefore corresponds to $\eta_0\simeq10$ mPa.s in the experiments. The capillary number $C_a$ compares the flow stress $\eta_0\dot{\gamma}$ with the elastic stress: $C_a=\eta_0\dot{\gamma} R /\mu$, where $R$ is the typical size of the RBC and $\mu$  the shear elasticity modulus of the membrane. $\eta_0$ is typically varied between 1 and 100 mPa.s and the stress $ \eta_0 \dot{\gamma}$ does not exceed 5 Pa (20 Pa in the more recent \cite{mauer18}). Technical details and description of the main results of the most recent and relevant papers that we consider here are reported in the Supplemental Material and will be recalled when necessary in the discussion of our results.

All recent simulation papers establish a diagram that is qualitatively coherent with the one partly drawn by experiments \cite[see][]{cordasco13,peng14,cordasco14_2,sinha15,mendez18}.  This diagram can be divided into four zones, the frontier of which depends on the considered equilibrium shapes (see Fig. \ref{fig:basicdiagram}): (i) flipping motions, (ii) tank-treading,  (ii') tank-treading like motions with oscillations or slightly off-plane motion, (iii) flipping motions, but  orbit stability is different than in the low external viscosity case (i). The intermediate region (ii') corresponds to a narrow range in the parameter space. In all papers, the different regimes are obtained by starting with a cell with a given orientation $\theta$, and following its time evolution. From the way the simulations are run, we can consider that they correspond more to the increasing $\dot{\gamma}$ case (but to a sharp increase). None of the previous studies consider the decreasing case, or the smoothly increasing case.

\subsection{Experimental papers}

One of the first major contributions to the study of RBC dynamics under shear flow is found in \cite{goldsmith72} where statistics on the angle distribution is obtained. They observed dilute red blood cells in a large tube where, at the cell scale, the flow can be assimilated to simple shear flow, with the same outer viscosity $\eta_0$ as plasma. They also performed  experiments in a plate plate geometry at high $\eta_0$ (520 mPa.s) and low $\dot{\gamma}$ (around 1 s$^{-1}$). Both experiments correspond to an increasing $\dot{\gamma}$: in the Poiseuille flow experiment the observation tube is narrower than the upstream tubes, and in the shear chamber experiment, RBCs are initially at rest. 

The main outcomes of Goldsmith \& Marlow study are : 
\begin{itemize}
\item In plasma and at low shear rate ($\dot{\gamma}<10$ s$^{-1},\,\tau<0.01$ Pa), the time evolution $\phi(t)$ can be well described by Jeffery's equation (Eq. \ref{eq:phi}) if an effective aspect ratio is introduced, which is 25\% lower than the aspect ratio of the convex envelope of the cell. This result was previously found for rigid circular disks (\cite{Anczurowski67}), and the discrepancy between the two ratios is expected to decrease as the aspect ratio gets closer to 1. 
\item At higher $\dot{\gamma}$ , departure from Jeffery's equation is observed, with longer duration of alignment with flow and  the initial $\phi\to-\phi$ symmetry for $\dot{\phi}$ is lost. This is interpreted as the consequence of cell deformability, which induces different shapes depending on the cell orientation relatively to the extensional direction of the flow.  At the same time, more cells drift into rolling motion  ($\theta=0$). Those feature are lost for hardened cells.
\item $\theta$ was found not to be periodic as a function of $\phi$, contrary to what the Jeffery theory states, and the motion appears to be less regular. This is interpreted as a consequence of rotary Brownian motion of cells.
\item In the highly viscous suspending medium, cells are seen to align strongly with the flow. However, no clear discrimination is made between tank-treading cells and rolling ones.\\
\end{itemize}

The drift to rolling is studied in \cite{bitbol86}. The cells were observed in a cone-plate geometry allowing for $\dot{\gamma}$  between 1 and 200 s$^{-1}$. Viscosity $\eta_0$ of the suspending medium was between 1 and 10 mPa.s. The main results are:
\begin{itemize}
\item If one considers the single parameter $\tau=\eta_0 \dot{\gamma}$, which measures the flow stress on the cell, a rough phase diagram is obtained from  direct observations as follows: if $\tau<0.02$ Pa, cells rotate in the flow (that is, flipping motion is observed). In the intermediate range 0.02 Pa $<\tau<1$ Pa, cells drift to rolling. For higher stresses, the number of rolling cells  plateaus at viscosity $\eta_0=1$ mPa.s while it decreases when $\eta_0>5$ mPa.s: cells start to tank-tread.
\item In the rolling regime, the cell diameter increases with $\eta_0$ and $\dot{\gamma}$, as already found in \cite{goldsmith72}.
\item In the considered viscosity range, the typical time for cell drift towards rolling  is of order 100$\times\dot{\gamma}^{-1}$.\\
\end{itemize}

Several new features were highlighted in \cite{dupire12}, which followed a first paper by the same group \cite[see][]{abkarian07}.
In these papers, the case of decreasing $\dot{\gamma}$  is considered for the first time, and off shear plane motion is also described for the first time. The carrying fluid viscosities $\eta_0$  are 7 and 29 mPa.s, and the $\dot{\gamma}$  goes from 0 to 15 s$^{-1}$ in the first case and to 2.7 s$^{-1}$ in the second case.  The cells flow in a large parallelepiped flow chamber where around 20 cells were tracked individually and visualized along both the shear  and  vorticity axis. 
The main results of this paper are : 
\begin{itemize}
\item At low $\dot{\gamma}$ , the orbit angle $\theta_0$ of a flipping cell is not constant but varies between 50$^{\circ}$ and 90$^{\circ}$. 
\item If the shear rate exceeds a threshold  $\dot{\gamma}_t$ lying between 0.01 Pa and 0.05 Pa depending on the observed cell, this orbit angle stabilizes to a constant value (that decreases with $\dot{\gamma}$ , until the rolling regime is reached). 
\item In the flipping regime, the time evolution of $\phi$ is well described by the Jeffery equation.
\item At some critical shear rate $\dot{\gamma}_c^+$, transition towards tank-treading is observed. Tank-treading is associated with swinging, that is, periodic oscillation of the inclination angle.
\item If $\dot{\gamma}$  is then decreased, the tank-treading motion remains stable until some other critical shear rate $\dot{\gamma}_c^-<\dot{\gamma}_c^+$ at which a transient intermittent regime is observed. It shows an alternance (in time) between tank-treading and flipping motions with high $\theta_0$. When an orbit angle lower than 50$^{\circ}$ is reached, flipping motion with fixed orbit angle becomes stable, with the same orbit angle equal as the one observed for the same shear rate in the increasing $\dot{\gamma}$ case. From this observation we deduce that $\dot{\gamma}_t<\dot{\gamma}_c^-$. Although not explicitly stated in \cite{dupire12}, it seems that upon a further decrease of $\dot{\gamma}$ the same branch as in the increasing $\dot{\gamma}$ case is followed.
\item $\dot{\gamma}_c^-$ and $\dot{\gamma}_c^+$ do not vary if many shear rate cycles are applied to the same cell. The corresponding critical shear stress values are 0.023 and 0.086 Pa for  $\eta_0=29$ mPa.s.  These are mean values, because all the critical stresses, as well as the orbit angles associated with a given stress, depend on the considered cell.
\item All these transitions occur in a shape preserving manner, with diameter variations that remains lower than 10\%. \\\end{itemize}

The transition between flipping and tank-treading for several viscosities and many cells is studied in \cite{fischer13}. They consider a Poiseuille flow, so probably only the increasing $\dot{\gamma}$ case is considered, though we do not know what the upstream conditions are. The critical $\dot{\gamma}$ for $\eta_0=11,\,24$ and 104 mPa.s are 75, 10 and 0.6 s$^{-1}$ respectively. The transition shear rate is determined as the value at which half of the red blood cells are in tank-treading regime. The authors also establish that the transition is not sharp: the number of cells in tank-treading increases smoothly with the $\dot{\gamma}$. The transition becomes more abrupt at high carrying fluid viscosity $\eta_0$. Those thresholds together with the one measured in \cite{morris79} for plasma viscosity, allow to draw a separation line between flipping-like motion and tank-treading-like motion in the ($\eta_0,\tau$) space, as shown in Fig. \ref{fig:basicdiagram}.\\

In \cite{lanotte16}, experiments similar to those in   \cite{fischer13} are performed in physiological condition, but on a limited amount of cells. Transitions from flipping to rolling are observed from $\dot{\gamma}=10$ s$^{-1}$ until $\dot{\gamma}=40$ s$^{-1}$ ($\tau=0.04$ Pa.s). In the meantime and also for larger shear rates, an increasing proportion of cup-shaped stomatocytes in rolling motion is detected. At even higher shear rates (until the maximum considered,  $\dot{\gamma}=2000$ s$^{-1}$), highly deformed polylobed shapes are observed. 
No notion of increasing or decreasing $\dot{\gamma}$  is introduced in these experiments. They confirmed this picture in a second recent paper with shear rates  higher than 60 s$^{-1}$ \cite[see][]{mauer18} \\

Finally, in \cite{levant16}, 17 cells are studied in a four roll mill apparatus which allows to consider a more general flow defined by the vorticity $\omega$ and the strain rate $s$. Simple shear flow corresponds to the case $2 \omega=2s =\dot{\gamma}$. It corresponds to a stability threshold for the apparatus, since $\omega/s>1$  is required for the cell to remain trapped. Long time observation is then possible, while varying the ratio $\omega/s$. More precisely, $\omega$ is fixed and $s$ is varied. The explored range of external viscosities is 20 to 87 mPa.s. The possibility to increase the contribution of the vorticity allows to switch to flipping regimes even in that viscosity range.  With the choice to characterize the flow stress by $2s \eta_0$ and to consider the extended viscosity $\eta_0 (w/s)^{-1}$, three regions are identified, corresponding to tumbling, swinging, and an intermittent regime, which coincide with those identified in \cite{dupire12}.  Contrary to what was observed in  \cite{dupire12},  large deformations are associated with the intermittent regime. Although the stress seems to be increased and decreased within the same experiment, no hysteretic behavior is reported. In Fig. \ref{fig:basicdiagram}, the explored area in the $(\eta_0 (w/s)^{-1},2s \eta_0)$ parameter space is shown on the phase diagram for simple shear rate with parameter space $(\eta_0,\eta_0 \dot{\gamma})$.  We shall note that, in agreement with numerical simulations in \cite{cordasco13}, no off-plane motion is observed in the flipping region while it is seen in simple shear flow in \cite{dupire12}. This partial mapping shows that an increased rotational component in the flow, as in \cite{levant16}, favors in plane motion.\\

The  experimental studies summarized above have all brought new features to the problem. They neither contradict each other nor provide cross-validation since the explored parameter ranges, or the experimental method, often differ.  Studies on individual cells which are followed in time have allowed to exhibit more detailed dynamics \cite[see][]{abkarian07, dupire12,levant16}, but  full characterization of the transition dynamics could not be addressed, in particular concerning the consequence of dispersion in size and mechanical properties of cells. The decreasing $\dot{\gamma}$  case has only been clearly addressed in \cite{dupire12}. On the other hand studies on statistically relevant populations are often limited to the study of a single feature, e.g. the population in tank-treading regime \cite[see][]{goldsmith72,fischer13} or in deformed rolling regime \cite[see][]{mauer18}, but they allowed to estimate the width of transition zones due to cell dispersity, in the increasing $\dot{\gamma}$  case only.

The case of physiological values for the external fluid viscosity has seldom been addressed, probably because of sedimentation issues, and is therefore much less documented than the case of more viscous fluids.

\subsection{On the hysteresis and the intermittent regimes}

A striking feature is the existence, at least for high enough viscosities of the external fluid, of an hysteresis in the $(\eta_0,\dot{\gamma})$ space upon variations of $\dot{\gamma}$, which has been described differently in different papers. We aim at clarifying this in the following.

In the first experimental paper to account for such a feature \cite[see][]{abkarian07}, the observed low shear rate motion is surprisingly always tumbling. Transition to tank-treading (or, rather, swinging) is characterized by the existence of an intermediate regime where cells alternatively tumble and swing, which is observed for both decreasing and increasing $\dot{\gamma}$. The experimental constraints did not allow to  conclude whether this intermittent regime is transient or not. Transition shear rates denoted $\dot{\gamma}_c^>$ and $\dot{\gamma}_c^<$ are identified, with $\dot{\gamma}_c^> >\dot{\gamma}_c^<$, a signature of hysteretic behavior. They correspond respectively to the first observed transition from tumbling to swinging when increasing $\dot{\gamma}$  and to the first observed transition from swinging to tumbling when decreasing $\dot{\gamma}$, for an observation over a time scale of order 20s.

In the same paper, an improvement of the analytical  model by Keller and Skalak\cite[see][]{keller82} (a droplet  enclosed by an ellipsoidal fluid membrane) is proposed for red blood cell dynamics. Membrane shear elasticity is taken into account  (as, simultaneously, in \cite{skotheim07}) and in \cite{dupire15}, the possibility for a stress free configuration different from  the equilibrium shape is explored. In that model, considering more inflated stress free configurations amounts to multiply the shear modulus by some constant smaller than 1. Considering the transition threshold and other characteristics like cell oscillation periods, it is deduced  in \cite{dupire15} that the stress free configuration is likely to be a spheroid, in agreement with other modelling \cite[see][]{lim02,peng14,cordasco14}. Back to our initial question, the analytical model results in differential equations for the angles characterizing cell dynamics, for the case where the cell axis of symmetry remains in the shear plane. In other words, only transitions between tumbling-like and tank-treading like motions can be explored (a recent model proposed in \cite{mendez18} should allow to get rid of this issue in the future). These equations can be solved numerically for a given shear rate and a given set of parameters for the cell mechanical properties, and an intermittent regime  is observed in an interval $[\dot{\gamma}_c^- ;\dot{\gamma}_c^+]$. In \cite{abkarian07} or in \cite{dupire15} it is however not specified if the threshold values are obtained considering increasing or decreasing $\dot{\gamma}$. Knowing the dependency of the solutions of the  cell angle  evolution equations  on the history of $\dot{\gamma}$ values  probably requires  a precise stability analysis such as that done in \cite{kessler09} in the case of quasi-spherical cells. 

Though they are clearly related to each other, how the different thresholds $\dot{\gamma}_c^{<,>}$  and $\dot{\gamma}_c^{-,+}$ compare with each other is not clear as they were determined differently (limited observation time in the experiments and more importantly no direct study of potential hysteresis in the numerical analysis).

The intermittent behavior has been observed and thoroughly studied by numerical simulations (starting from a rest configuration) in \cite{cordasco14} and in \cite{peng14}, with the axis of revolution also restricted to shear plane, though in some cases this configuration is said to be metastable according to \cite{cordasco14}. 
In \cite{peng14}, the minor bound $\dot{\gamma}_c^-$ of the intermittent regime is determined 
and  shown to be smaller than the transition shear rate between rolling and tank-treading. Finally, in the experimental paper by \cite{levant16}  a stationary intermittent regime is also observed for, apparently, increasing and decreasing flow stress although it is only explicitly highlighted for a decrease of the flow stress $2 s \eta_0$. 
Note that in this paper the flow geometry is not the same and no off-shear plane motion is observed.

In \cite{dupire12}, the picture turns out to be more complex: the decreasing $\dot{\gamma}$ path would be characterized by the apparition of a transient intermittent (tank-treading / flipping with angle larger than 50$^\circ$) phase when tank-treading regime is left, whose duration in time is not clear. After this intermittent phase, flipping motion with orbit angle strictly smaller than 50$^\circ$ would follow. The shear rate at which this transition takes place is denoted $\dot{\gamma}_c^-$, and no notion of interval over which this intermittent regime would be possible is introduced. The analogy with the notation  $\dot{\gamma}_c^-$ introduced in \cite{abkarian07} suggests that the transient intermittent regime would only take place in the low range of the whole expected range for the  intermittent regime $[\dot{\gamma}_c^- ;\dot{\gamma}_c^+]$, while stable tank-treading would be preferred in the top range. 

On the other hand, upon an increase of shear rate, no intermittent regime is observed but rather a continuous drift from tumbling to rolling, then a transition to swinging. In \cite{dupire12}, the threshold for this transition is denoted $\dot{\gamma}_c^+$, suggesting that when increasing $\dot{\gamma}$  the route to swinging through the intermittent regime is replaced by the orbital drift. In \cite{levant16}, the intermittent regime is associated with large cell deformations: we hypothesize that the orbital drift is another route that prevents such large elastic distortions.  This would explain why the intermittent regime is not seen in the increasing shear rate case, unless forced by restricting the motion to the shear plane, artificially in the simulations or because of a specific flow geometry as in \cite{levant16}.

Finally, the fact that, to our knowledge, no analytical or numerical analysis has considered the decreasing shear rate case so far limits the discussion on the relationship between hysteretic and (possibly transient) intermittent regimes. A secondary goal of the present study is to reinforce the experimental input on this question.

\section{Experiments}

Blood samples from  healthy donors were tested and provided by the French blood bank EFS. To remove all substances except RBCs, the samples were centrifuged and washed in physiological buffer solution (PBS, phosphate buffered saline, 
Sigma). This procedure was repeated three times at room temperature; after each centrifugation, the liquid phase and buffy coat were removed by aspiration. Then  RBCs were resuspended either in PBS solution with 1 g/L of bovine serum albumine (BSA, Sigma)  or in a PBS+BSA solution where dextran was diluted (20 g/L of dextran of molecular weight 1.5$\times 10^4$ + 70 g/L of dextran of molecular weight 2$\times 10^6$). This latter solution has a viscosity at room temperature (23-25 $^\circ$ C) of  25$\pm 0.5$ mPa.s, a configuration close to that often considered in the literature. The viscosity and density of this solution are such that they allow long-time acquisition without sedimentation issues.

By contrast, cells in plasma quickly sediment, with a velocity of typically 1 \textmu m/s. This prevents long-time observations in flow chambers adapted to microscopy. In order to counterbalance sedimentation, water in the PBS solution was replaced by a mixture of 68.5 \% water and 31.5 \% Optiprep$^{\copyright}$ (iodixanol solution, Axis-Shield), so as to reach a density close to that of cells (1.1 g/ml), following a method initially proposed in \cite{roman12}, later on used in \cite{shen16}Ê and \cite{roman16}. This solution has a viscosity at room temperature of 1.5$\pm0.1$ mPa.s, close to that of plasma (1.95 mPa.s at 20$^\circ$C and 1.34 mPa.s at 37$^\circ$C \cite[see][]{brust13}). 

\begin{figure}
\begin{center}
  \includegraphics[width=\columnwidth]{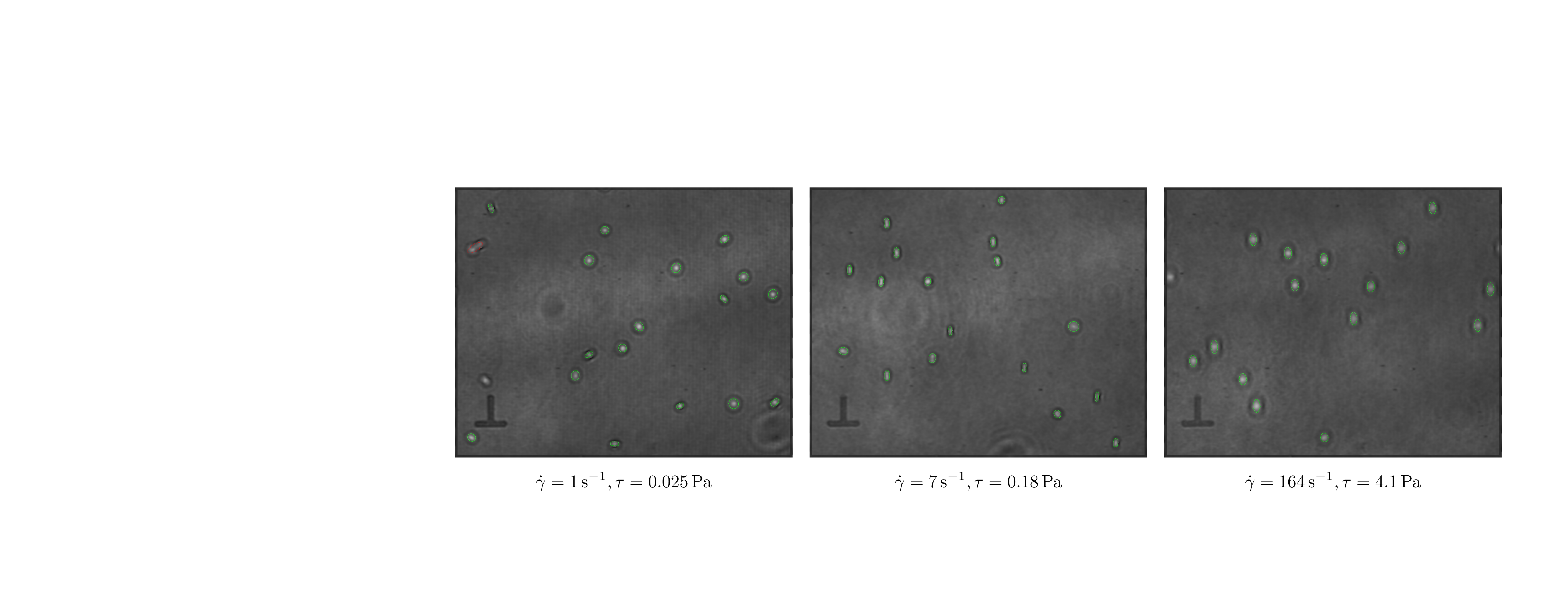}
\caption{Examples of images from the DHM after treatment. The green contours are the fits of the cells projected shapes. Red ones correspond to contours that were discarded because the aspect ratio or the apparent surface are considered as non realistic. They mostly correspond to superimposed cells. Due to the low concentration in cells, those patterns are scarce and, whatever the exclusion criterion considered, do not influence much the analysis.\label{fig:exDHM}}
\end{center}
\end{figure}

In all suspensions, the volume fraction is very weak ($\sim 0.01\%$), and cells are located everywhere in the chamber: one can consider the cells are isolated and do not interact hydrodynamically. The suspension is injected in a plane-plane shear chamber of gap $200$ \textmu m which was previously described in \cite{callens08}. The shear rate $\dot{\gamma}$ is varied step by step from 0.5 s$^{-1}$ to 200 s$^{-1}$ then decreased back.
The flow stress ranges are thus $7.5\times10^{-4}$ Pa to $ 0.3$ Pa  and $1.3 \times 10^{-2}$ Pa to 5 Pa for the low and high viscosity solutions, respectively. Those ranges are indicated in Fig. \ref{fig:basicdiagram} for comparison with the explored ranges in the literature. Because of the low sedimentation rate, one sequence of increasing or decreasing shear rate could last around 1h, with the same RBCs in the chamber all along the experiment. Thus, small steps for shear rate variation could be imposed, so as to obtain a precise diagram of the RBC dynamics.  At each value of $\dot{\gamma}$, between 1 and 5 (for low shear rates) series of 600 images are taken at 24 fps. Series are separated by at least 15 s. Using several series allows obtaining good enough statistics on the shape and orientation of the cells.  Only cells located at at least 20 \textmu m from the walls are included in the analysis, because the presence of walls may impact the cell dynamics, in particular flipping motion \cite[see][]{vitkova09}.

    The RBC suspension is monitored by a Digital Holographic Microscope working in reduced spatial coherence, whose optical axis is perpendicular to the chamber planes. This configuration allows a strong reduction of the inherent noise of interferometry and provides better in-depth reconstruction capabilities \cite[see][]{grandchamp13,minetti16}. On the basis of the acquired interferogram, it is possible to extract the complex amplitude of the optical beam passing through the sample and simulate numerically the propagation of the beam through the whole thickness of the experimental cell. In this way one can scan, slide by slide, the experimental volume and refocus numerically all the RBCs present in the suspension volume.  The procedure to extract the three-dimensional position and features of each cell in the experimental volume is similar to the one used in \cite{minetti14} for lipid vesicles. Red blood cells require more attention in the data treatment than vesicles because their convex shape leads to a much higher distortion of the light beam, a property which was recently used to characterize cell morphologies \cite[see][]{miccio15}. The projection of the cell shape on the plane perpendicular to the shear gradient direction is obtained, where each pixel intensity is proportional to the optical thickness (thickness of the cell times the difference of refractive index between the cell and the suspending fluid) in the optical axis direction. A segmentation is applied on each cell to determine the geometrical boundaries of the projected shape.

For each detected cell, its major axis is determined from the diagonalisation of the inertia matrix of its projected shape. This axis makes an angle $\Psi$ with respect to the flow direction, as defined is Sec. \ref{sec:euler}. $\Psi$ is defined in the range $[0,90^\circ]$. Taking this axis as an origin for angle $\xi$, the contour of the projection of the cell is fitted by the polar equation $\rho(\xi)=\rho_0 + a_1 \cos^2\xi + a_2 \cos^4\xi+ a_3 \cos^6\xi$, which allows to also describe shapes that are more rod-like than ellipse-like, as is the case when viewed from the side. In that case, they may also appear as concave, because of the smaller contribution of the outer ring to the total optical thickness of the object. Examples of obtained images with the associated contours are shown in Fig. \ref{fig:exDHM}. The weight of the contribution of the outer ring can vary a lot depending on the orientation or position of the cell, for that reason we consider the aspect ratio $r_a$ of the convex envelope of the cell (that is, its effective projection) rather than the ratio between the  radii at $\xi=0$ and $\pi/2$ : $r_a=\rho(0)/\max\big( \rho(\xi) \sin \xi\big )$. $r_a$ lies in the range $[1;+\infty]$, but in practice it is never larger than 5. Probability densities for $\Psi$ and $r_a$ are then calculated for each shear rate. For simplicity we consider the $\Psi$ and $r_a$ distributions separately, so as to make the fitting procedure described in the next section converge more easily and in a reasonable amount of time. A long computing time is due to the necessity to consider only discrete data sets instead of formal expressions for some fitting functions, as described in the next section. The bin sizes are $\delta r_a=0.05$ for  $r_a$  and  $\delta \Psi=1^\circ$ for $\Psi$. For high $\dot{\gamma}$, around 6000 cell pictures are included in the statistics. This number is regularly increased as lower shear rates are considered, up to 40 000 for the lower shear rate 0.5 s$^{-1}$, so as to  describe accurately more complex dynamics with many orbit angles. This last number corresponds to about 500 different cells that were observed at least once.


\section{Model and data analysis}

The experimental probability densities are compared to the theoretical distribution  for $\Psi$ and $r_a$  that are expected from the model described in the following.

\subsection{Premises}

We assume that, for a given shear rate, the suspension is composed at each time, on average, of a proportion $p_{T}$ of cells  in tank-treading-like (or swinging) motion and a proportion $1-p_{T}$ of cells in flipping motion following a Jeffery orbit, where orbit angle $\theta_0$ can be between 0 (rolling) and $90^\circ$ (tumbling). In \cite{dupire12}, it is shown that above a threshold value, orbits are unstable and the cells switch from one orbit to another, with angles between this threshold and $90^\circ$. On the contrary, below this value orbits are stable. In order to explore this idea, we look for two subpopulations of flipping cells: a) with proportion $p_s$, those with stable orbits which we suppose to lie between two extremal values $\theta^-_0$ and $\theta_0^+$, with equal probability. While this may appear as a strong statement, this is the only reasonable choice amongst the possible distributions, that would lead to a reasonable number of fitting parameters. b) with proportion $(1-p_s)$, those with unstable orbits with angles between $\theta^+_0$ and $90^\circ$, with equal probability. In addition, shear stress may have an influence on the cells' aspect ratio, which should also be determined.

In the following, we derive the expression of the theoretical densities that include the geometrical and dynamical characteristics of the tank-treading and flipping regimes, but also the dispersion in size and shape within a sample and the  uncertainties associated with the  shape characterizations.


\subsection{Tank-treading}


\begin{figure}
\begin{center}
  \includegraphics[width=\columnwidth]{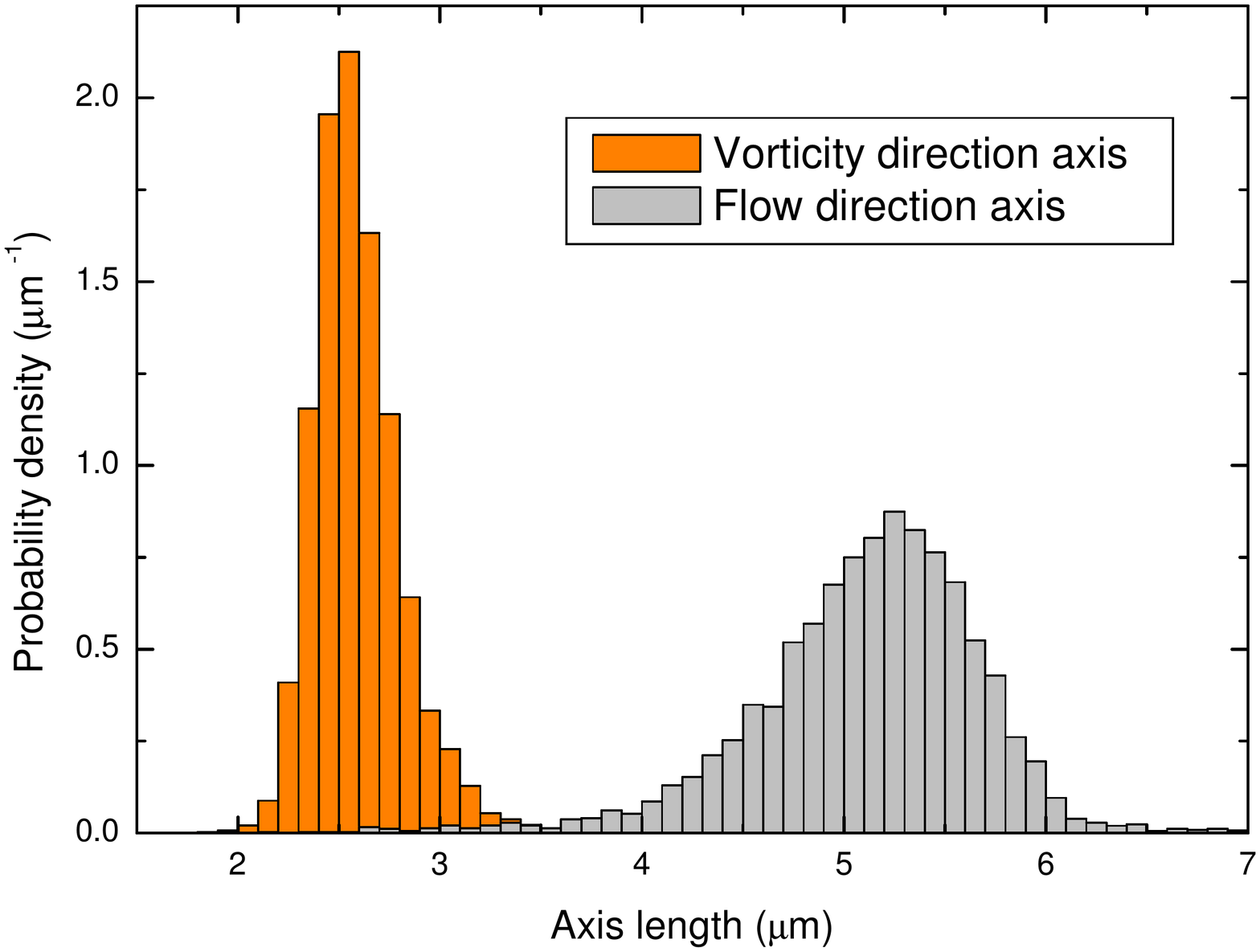}
\caption{Probability density for the length of the projected shapes of RBCs, in the vorticity and flow direction, in the $\eta=25$ mPa.s solution, for $\dot{\gamma}=200$ s$^{-1}$.\label{fig:axis}}
\end{center}
\end{figure}

Viewed from the shear axis direction, depending on the elongation of the cell (increasing with shear rate), but also on its angle $\pi/2-\phi$ compared to the flow direction (decreasing with shear rate), a tank-treading (or swinging)  cell will appear as an ellipse whose long axis may be in the flow or in the vorticity direction. 
We therefore look for two populations of cells, in proportion $\beta$ and $1-\beta$, with orientations 0 and $90^\circ$. The theoretical distribution $d_{\Psi,TT}$ on $[0^\circ;90^\circ]$ is then, in principle: \begin{equation}d_{\Psi,TT}[\beta](\Psi)=\beta\, 2\, \delta_0(\Psi)+(1-\beta)\,2 \,\delta _{90}(\Psi),\label{distangTTth}\end{equation} where $\delta_{\Psi_0}$ is the Dirac distribution centered on $\Psi_0$.

Here and in the following, we mention inside square brackets $[-]$ the  parameters  of the distribution while the studied variable is placed inside brackets $(-)$. For the aspect ratio, we notice that the size distribution is large, as at rest: typical range for  the long axis of a cell at rest is 7 to 9 \textmu m \cite[see][]{canham68}. In addition, the distribution of apparent cell length in the flow direction is not symmetric but has shorter tail on the large values side (see Fig. \ref{fig:axis}). The distribution of width in the vorticity direction is not symmetric either, with shorter tail on the low value side. This may be due to many factors: uneven distribution of the deformability among the cells \cite[see][]{Dobbe02_1}, saturation of deformation of a given cell under shear (it would not possible to elongate a cell and shrink it in the vorticity direction beyond a given threshold), result of swinging dynamics that favors some  projected lengths (not valid for width). For the aspect ratio $r_{zx}=$(flow direction axis / vorticity direction axis) we shall therefore expect an asymmetric distribution \cite[see][]{Dobbe02_2}. It appeared that the skew-normal distribution, defined by \begin{equation} \label{eq:skew} p_{sG}[r_T,w_T,\alpha](r_{zx})\propto (1+erf(\scriptstyle \frac{\alpha (r_{zx}-r_T)}{\sqrt{2} w_T}\displaystyle))e^{-(r_{zx}-r_T)^2/(2 w_T^2)}\end{equation} works well. $\alpha$ is an asymmetry parameter that will be negative in our case. In our general analysis, we chose  to consider the aspect ratio $r_a$ between the long axis and the short axis of the cells, a parameter that is defined even for cells not parallel or perpendicular to flow.  Here, we shall therefore consider the inverse aspect ratio for the cells with aspect ratio $r_{zx}$ lower than 1, hence the following theoretical distribution $d_{r_a,TT}$ for $r_a$ :\begin{equation} 
d_{r_a,TT}[r_T,w_T,\alpha](r_a)=\Theta (r_a-1) \Big[ p_{sG}[r_T,w_T,\alpha](r_a)+ \frac{\scriptstyle 1}{\scriptstyle r_a^2}  p_{sG}[r_T,w_T,\alpha](\frac{\scriptstyle 1}{\scriptstyle r_a} )\Big], \label{eq:raTT}
\end{equation}
with $\Theta$ the Heaviside function. 
The proportion $\beta$ of cells whose projection is aligned with the flow (Eq. \ref{distangTTth}) , is equal to $\int_1^{\infty}  p_{sG}[r_T,w_T,\alpha](r_a) dr_a$.

\subsection{Flipping motion}

Although cells are not exactly ellipsoids, we assume that the dynamics of a flipping cell may still be described by the Jeffery orbit of an ellipsoid having the same aspect ratio $r$ as the convex envelope of the cell, as in \cite{Anczurowski67}.

In experiments, the optical axis is $y'$ and we visualize its projection on the $x'z'$ plane (see Fig. \ref{fig:schema}  for the notations and Fig. \ref{fig:exJeff}  for an example). 
The relationship between the coordinates $(x,y,z)$ in the moving $Oxyz$ coordinate system and  $(x',y',z')$ in the $Ox'y'z'$  coordinate system is, according to the convention used here: \begin{equation}\label{eqrot}\begin{pmatrix}x\\y\\z\end{pmatrix}
=\Big[ \begin{pmatrix}
   1 & 0 &0 \\
   0 & \cos \phi & - \sin \phi \\
   0 & \sin \phi & \cos \phi\\ 
\end{pmatrix}\begin{pmatrix}
   \cos \theta & - \sin \theta & 0 \\
 \sin \theta& \cos \theta &0\\ 
 0&0&1\\
\end{pmatrix}\Big]^{-1}\begin{pmatrix}x'\\y'\\z'\end{pmatrix}.\end{equation}

Eqs \ref{eqellips} and \ref{eqrot} yield the equation of the ellipsoid in the $Ox'y'z'$ coordinate system. This is a quadratic equation in $x',y'$ and $z'$, with parameters $r$, $\theta$ and $\phi$. The contour of the projection of the ellipsoid along the $y'$ axis corresponds to the $x'z'$ points for which the ellipsoid equation has exactly one solution for $y'$. It yields the quadratic equation of an ellipse for  $x'$ and $z'$ which can be written as $A(r,\theta,\phi)x'^2+ B(r,\theta,\phi)x'z'+C(r,\theta,\phi)z'^2=1$.

The angle $\Psi$ of the long axis relatively to flow direction $z'$ and the aspect ratio (long axis/short axis) $r_{a}$ of this ellipse are given by  \begin{eqnarray} \cot \Psi &=&(- A  + C - \sqrt{(A - C)^2 + B^2})/B, \\ r_a&=&\sqrt{(A + C + \sqrt{(A - C)^2 + B^2})/(A+ C - \sqrt{(A - C)^2 + B^2})}. \label{eq:rath}\end{eqnarray}

For a given aspect ratio $r$ and orbit parameter $\theta_0$ we can calculate the trajectories $\Psi(r,\theta_0,t)$ and $r_a(r,\theta_0,t)$  over one period and deduce the associated probability function for $\Psi$ and $r_a$, as the probability is inversely proportional to the time derivative of the considered parameter.  For $r_a$ it yields a very complex function, which furthermore has to be integrated to normalize the probability, making the calculation unfeasible. For $\Psi$, thanks to the more direct Eq. \ref{eq:PsiJeff}, this can indeed be more easily done, and one finds, for $\Psi$ cast onto $[0;90^{\circ}]$, the probability density: \begin{equation} p_{\Psi}[\theta_0](\Psi)=\frac{1 + \tan^2 \Psi}{90 \tan \theta_0 \sqrt{1 - \tan^2\Psi/\tan^2\theta_0}}\quad \mbox{   if } \Psi<\theta_0, \mbox{ and 0 else.} \end{equation}

Interestingly, for a given $\theta_0$, the apparent angle probability does not depend on $r$. This could already be seen from the expression \ref{eq:PsiJeff}, where the aspect ratio $r$ only appears in the expression through the period.

To calculate the probability density $p_{r_a}$  for $r_a$ we chose to sample $r_a(r,\theta_0,t)$ given by Eq. \ref{eq:rath} over one period $T$  with time step $10^{-3} T$ and infer the associated probabilities $p_{r_a}[r,\theta_0](r_a)$ with bin size 0.05 as for the experimental data,  for all aspect ratios $r$ between $r_{\min}=1.025$ and $r_{\max}=4.975$ with step $\delta r=0.05 $ and all orbit angles $\theta_0$ between  0 and $90^\circ$ with step $\delta \theta_0=0.25^\circ$.  This set of 80$\times$361=28880 discretized probability distributions will be used in the fitting procedure.

Following the premises of this modelling section, the expected distributions are eventually obtained  by considering that the orbit angle lies between two extremal values $\theta^-_0$ and $\theta_0^+$, with equal probabilities of sum $p_s$ or between $\theta^+_0$ and $90^\circ$, with equal probabilities of sum $(1-p_s)$ and that, because cell population is not fully homogeneous, there  can be some variation in their aspect ratio. We consider normal $p_G[r_{F},w_{F}](r)$ distribution for the cell aspect ratio \cite[see][]{rodak}, with mean $r_{F}$ and standard deviation $w_{F}$. We would thus seek for $\Psi $ and $r_a$  distribution in flipping regime \begin{multline}
d_{r_a,F}[r_{F},w_{F},\theta^-_0,\theta_0^+,p_s](r_a)=\sum_{r=r_{\min}}^{r_{\max}}p_G[r_{F},w_{F}](r) \delta r  \\
 \times \Big( \frac{\scriptstyle p_s \, \delta\theta_0}{\scriptstyle \theta^+_0-\theta^-_0+  \delta\theta_0}  \sum_{\theta_0=\theta_0^-}^{\theta_0^+}  p_{r_a}[r,\theta_0](r_a)+\frac{\scriptstyle (1-p_s)\, \delta\theta_0}{\scriptstyle 90-\theta^+_0 +\delta\theta_0}  \sum_{\theta_0=\theta_0^-}^{90}  p_{r_a}[r,\theta_0](r_a)\Big), \label{eq:raF}
 \end{multline}
 \begin{equation}
d_{\Psi,F}[\theta^-_0,\theta_0^+,p_s](\Psi)= \frac{\scriptstyle p_s \, \delta\theta_0}{\scriptstyle \theta^+_0-\theta^-_0+  \delta\theta_0} \sum_{\theta_0=\theta_0^-}^{\theta_0^+} p_{\Psi}[\theta_0](\Psi)+ \frac{\scriptstyle (1-p_s)\, \delta\theta_0}{\scriptstyle 90-\theta^+_0 +\delta\theta_0} \sum_{\theta_0=\theta_0^+}^{90} p_{\Psi}[\theta_0](\Psi).\label{eq:PsiF}
\end{equation}

Sums on $r$ are made with step $\delta r$ and sums on $\theta_0$ use step $\delta \theta_0$. Both steps are those used for the generation of the reference distributions.

\subsection{Error function on angles}

Experimental uncertainty on aspect ratio determination is somehow taken into account by the use of (skew-)normal distributions.

The determination of the cell angle may lead to huge errors when cell apparent aspect ratio is close to 1. If one considers the ellipse equation $A x^2 + B xz + C z^2 = 1$ with $A,B$ and $C$ being normal distributed around $1,0$ and $1+\epsilon$ respectively and variances of order 0.1 close to the experimental ones, one numerically finds an orientation angle distribution that is centered on 0 but with long tail such that it is rather well described by Cauchy distribution. As the considered angles are in the interval $[0;90]$ we will therefore consider, instead of $\delta$ functions, the Cauchy distribution folded onto this interval: \begin{equation}\tilde{p}_C[\Psi_0,w_{\Psi}](\Psi)=p_C[\Psi_0,w_{\Psi}](\Psi)+p_C[\Psi_0,w_{\Psi}](-\Psi)+p_C[\Psi_0,w_{\Psi}](180-\Psi),\end{equation}
with $p_C[\Psi_0,w_{\Psi}](\Psi)=\frac{1}{w_{\Psi} \pi (1 + \frac{\scriptstyle(\Psi-\Psi_0)^2}{\scriptstyle w_{\Psi} ^2})} $ the Cauchy distribution function around $\Psi_0$.

For the tank-treading regime, we will thus consider the modified distribution function for the angle : \begin{equation}
\hat{d}_{\Psi,TT}[\beta,w_{\Psi}](\Psi)=\beta \tilde{p}_C[0,w_{\Psi}](\Psi)+(1-\beta) \tilde{p}_C[90,w_{\Psi}](\Psi).\label{eqPsiTTmod}
\end{equation}

$w_{\Psi}$ is a free parameter that is expected to increase when $\dot{\gamma}$   decreases, to take into account the fact that the aspect ratio becomes close to 1.

For the flipping cells, the link between aspect ratio and angle is more complex. The apparent aspect ratio is maximal when $\Psi=\theta_0$ (cell viewed from the edge, $\phi=\pm90^\circ$). It is minimal when $\Psi=0$ but this minimal value depends on the orbit: it reaches 1 only for tumbling ($\theta_0=90^\circ$). Therefore, the uncertainty on a given angle depends on the orbit the cell is following. So as to go on handling separately the $r_a$ distribution and the $\Psi$ distribution, we make the simplifying assumption that, for a given orbit, the width of the modified Cauchy function is 0 when the angle is equal to its maximum value $\theta_0$ and increases linearly with the distance between the angle and the orbit angle. We thus consider the modified distribution for one orbit: \begin{equation}
 \hat{p}_{\Psi}[\theta_0,w_{\Psi}](\Psi)= \sum_{\xi=0}^{90} p_{\Psi}[\theta_0](\xi) \tilde{p}_C[\xi,w_{\Psi}|\xi-\theta_0|/90](\Psi) .\end{equation}

Here, $w_{\Psi}$ is the width of the error distribution when the aspect ratio is close to 1 ($\theta_0=90^\circ, \xi=0$). For simplicity, we shall consider the same $w_{\Psi}$ for both tank-treading and flipping distributions. Both tank-treading and flipping populations will coexist in the highly viscous solution. At high shear rate, tank-treading cells have aspect ratio far from 1 so $w_{\Psi}$ will be smaller than the expected value for the flipping cells, but in that case, the latter are not many. As $\dot{\gamma}$  decreases, flipping cells will appear but in the meantime the apparent aspect ratio of tank-treading cells will be close to 1, and the $w_{\Psi}$ values should coincide reasonably.

\begin{table}
  \begin{center}

  \begin{tabular}{ll}
  
  $p_{T}$ & Proportion of TT cells \\
  $r_T$ & Location parameter of the skew-normal distribution for the apparent aspect ratio of TT cells  \\
  $w_T$ & Scale parameter of the skew-normal distribution for the  apparent aspect ratio of TT cells \\
  $\alpha$& Shape parameter of the skew-normal distribution for the apparent aspect ratio of TT cells\\
  $r_{F}$& Mean aspect ratio of the F cells, assimilated to oblate ellipsoids\\ 
    $w_{F}$& Standard deviation of the aspect ratio of the F cells, assimilated to oblate ellipsoids\\ 
    $\theta^-_0$& Minimal stable orbit angle of F cells\\
        $\theta^+_0$& Maximal stable orbit angle of F cells\\
        $p_s$ &Proportion among the F cells in stable orbits (between $\theta^-_0$ and $\theta^+_0$)\\
            $w_{\Psi}$& Width of the error distribution of apparent angles\\
  \end{tabular}
  \caption{Summary of the fitting parameters for the apparent angle and apparent aspect ratio distributions. See Eqs. \ref{eq:ra} and \ref{eq:Psi}. TT refers to tank-treading and F to flipping.}
  \label{tab:param}
  \end{center}
\end{table}

Finally, the modified distribution function for the flipping regime is 

\begin{eqnarray}
\hat{d}_{\Psi,F}[\theta^-_0,\theta_0^+,p_s,w_{\Psi}](\Psi)&=& \frac{\scriptstyle p_s \, \delta\theta_0}{\scriptstyle \theta^+_0-\theta^-_0+  \delta\theta_0} \sum_{\theta_0=\theta_0^-}^{\theta_0^+}  \hat{p}_{\Psi}[\theta_0,w_{\Psi}](\Psi)+ \frac{\scriptstyle (1-p_s)\, \delta\theta_0}{\scriptstyle 90-\theta^+_0 +\delta\theta_0} \sum_{\theta_0=\theta_0^+}^{90}  \hat{p}_{\Psi}[\theta_0,w_{\Psi}](\Psi)\nonumber\\
&&\label{eq:PsiFmod}
\end{eqnarray}

\subsection{Complete fitting function}

Considering that we have two populations of cells, one in tank-treading-like regime with proportion $p_{T}$ and one in flipping regime in proportion $(1-p_{T})$, we finally have the following fitting functions for the distributions in apparent aspect ratio $r_a$ and apparent angle $\Psi$, respectively defined on $[1;+\infty]$ and $[0;90]$, with 10 fitting parameters: 
 \begin{multline}
d_{r_a}[p_{T},r_T,w_T,\alpha,r_{F},w_{F},\theta^-_0,\theta_0^+,p_s](r_a)=\\
p_{T} \,\,d_{r_a,TT}[r_T,w_T,\alpha](r_a) +(1-p_{T})\,\,d_{r_a,F}[r_{F},w_{F},\theta^-_0,\theta_0^+,p_s](r_a), \label{eq:ra}
\end{multline}
with $d_{r_a,TT}$ given by Eq. \ref{eq:raTT} and $d_{r_a,F}$ by Eq. \ref{eq:raF}, and \begin{multline}
d_{\Psi}[p_{T},r_T,w_T,\alpha,\theta^-_0,\theta_0^+,p_s,w_{\psi}](\Psi)=\\
p_{T} \,\,\hat{d}_{\Psi,TT}[\beta,w_{\Psi}](\Psi)+(1-p_{T}) \,\,\hat{d}_{\Psi,F}[\theta^-_0,\theta_0^+,p_s,w_{\Psi}](\Psi),\\
\mbox{with } \, \beta = \int_1^{\infty}  p_{sG}[r_T,w_T,\alpha](r_a) dr_a.
\label{eq:Psi}
\end{multline}

$\hat{d}_{\Psi,TT}$ is given by Eq. \ref{eqPsiTTmod} and $\hat{d}_{\Psi,F}$ by Eq. \ref{eq:PsiFmod}.

The fitting parameters are recalled in Table \ref{tab:param}. For each shear rate, we minimize the distance  between the theoretical and experimental distribution functions for $r_a$ and $\Psi$, given by \begin{multline} \epsilon = \sum_{\Psi}  | d_{\Psi,exp} (\Psi)-d_{\Psi}[p_{T},r_T,w_T,\alpha,\theta^-_0,\theta_0^+,p_s,w_{\psi}](\Psi)| \delta \Psi\quad + \\ \sum_{r_a} | d_{r_a, exp}(r_a)-d_{r_a}[p_{T},r_T,w_T,\alpha,r_{F},w_{F},\theta^-_0,\theta_0^+,p_s](r_a)| \delta r_a. \label{eq:epsilon}\end{multline}

Because they appear as bounds in sums, the parameters $\theta_0^-$ and $\theta_0^+$ are treated differently from the other parameters. For a given choice of these angles, we minimize $\epsilon$ using the NMinimize function of Mathematica $^\copyright$ software. We then explore systematically the region of interest for these angles so as to find the global minimum.

\section{Results}

\subsection{Preliminary result: hardened cells}

\begin{figure}
\begin{center}
  \includegraphics[width=\columnwidth]{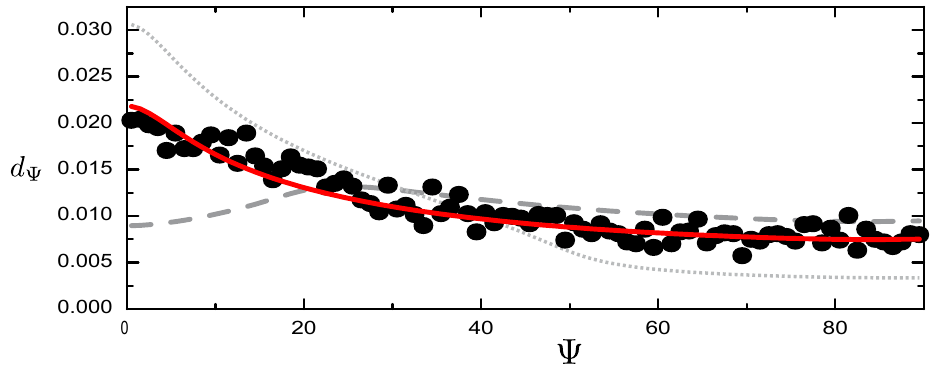}
\caption{Apparent angle $\Psi$ probability distribution for hardened cells in the suspending medium of viscosity 1.5 mPa.s, for different $\dot{\gamma}=200 $ s$^{-1}$. Dots: experimental data; red solid line: full fit (Eq. \ref{eq:Psi}) with $p_T=0$ (no tank-treading cells), $\theta_0^-=0^\circ$, $\theta_0^+=90^\circ$. In order to help understand more complex distributions, we also show the distribution for $\theta_0^-=20^\circ$, that exhibits a peak around $20^\circ$ (dotted line), and a case where the probability of orbits between $50^\circ$  and  $90^\circ$ is 1/4 of that between $0^\circ$  and  $50^\circ$  (dashed line). A strong decrease can be seen around $50^\circ$. \label{fig:rigides}}
\end{center}
\end{figure}

Glutaraldehyde-hardened cells were studied at $\dot{\gamma}=200$s$^{-1}$ in the  $\eta_0=1.5$ mPa.s solution. No tank-treading motion is expected and we find that the apparent angle distribution is fully compatible with that of flipping cells with orbit angles equally distributed in the $[0^\circ;90^\circ]$ interval, as shown in Fig. \ref{fig:rigides}. This is agreement with the fact that Jeffery orbits are stable at low Reynolds number: the followed orbit only depends on the initial condition.

\subsection{Cells in fluid of high viscosity}

\begin{figure}
\begin{center}
  \includegraphics[width=\columnwidth]{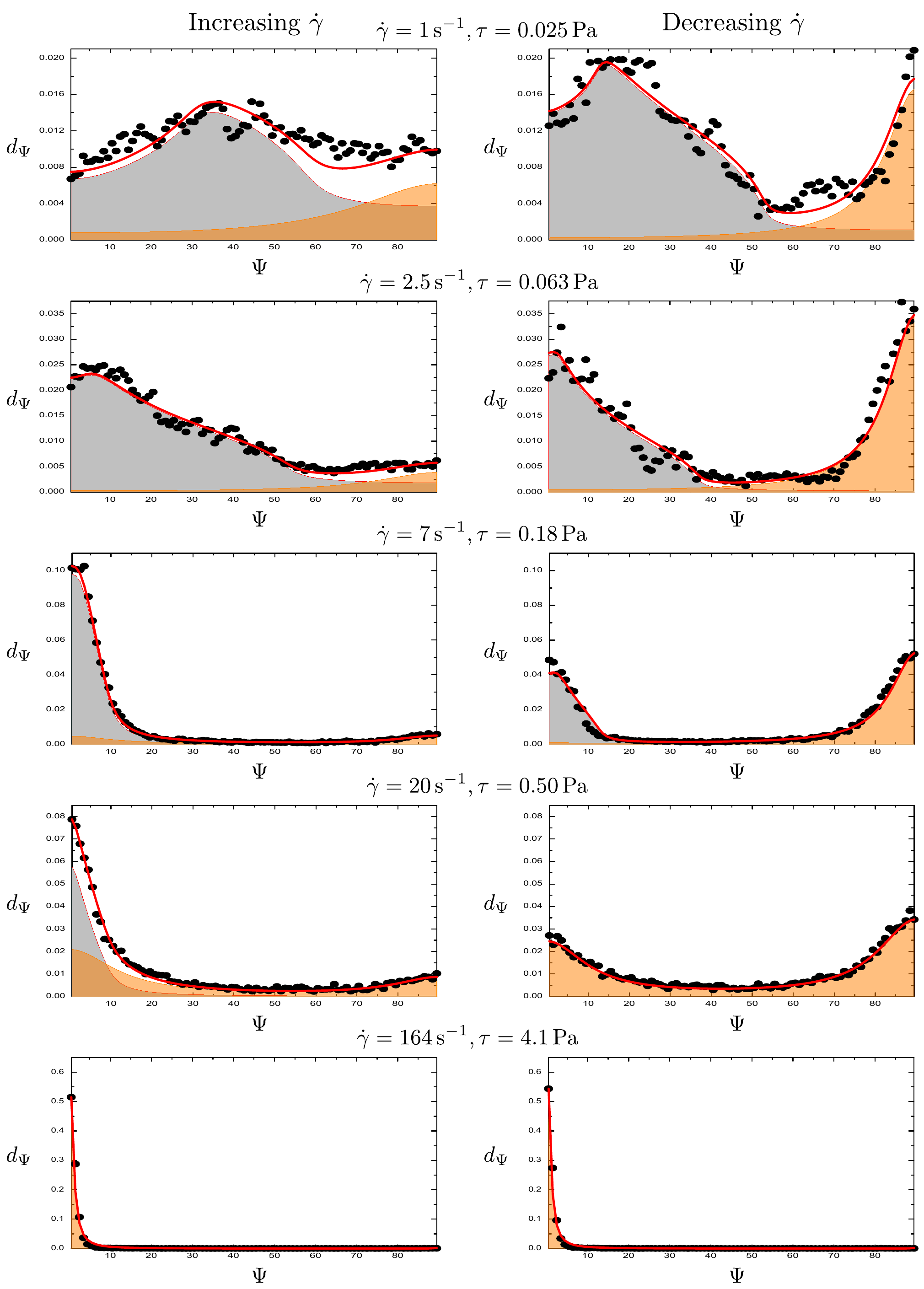}
\caption{Selection of apparent angle $\Psi$ probability distributions for cells in the suspending medium of viscosity 25 mPa.s, for different shear rates. Dots: experimental data; red line: full fit (Eq. \ref{eq:Psi}); orange line and area: tank-treading population; red line and grey area: flipping population.}\label{fig:exemples-angles}
\end{center}
\end{figure}

\begin{figure}
\begin{center}
  \includegraphics[width=\columnwidth]{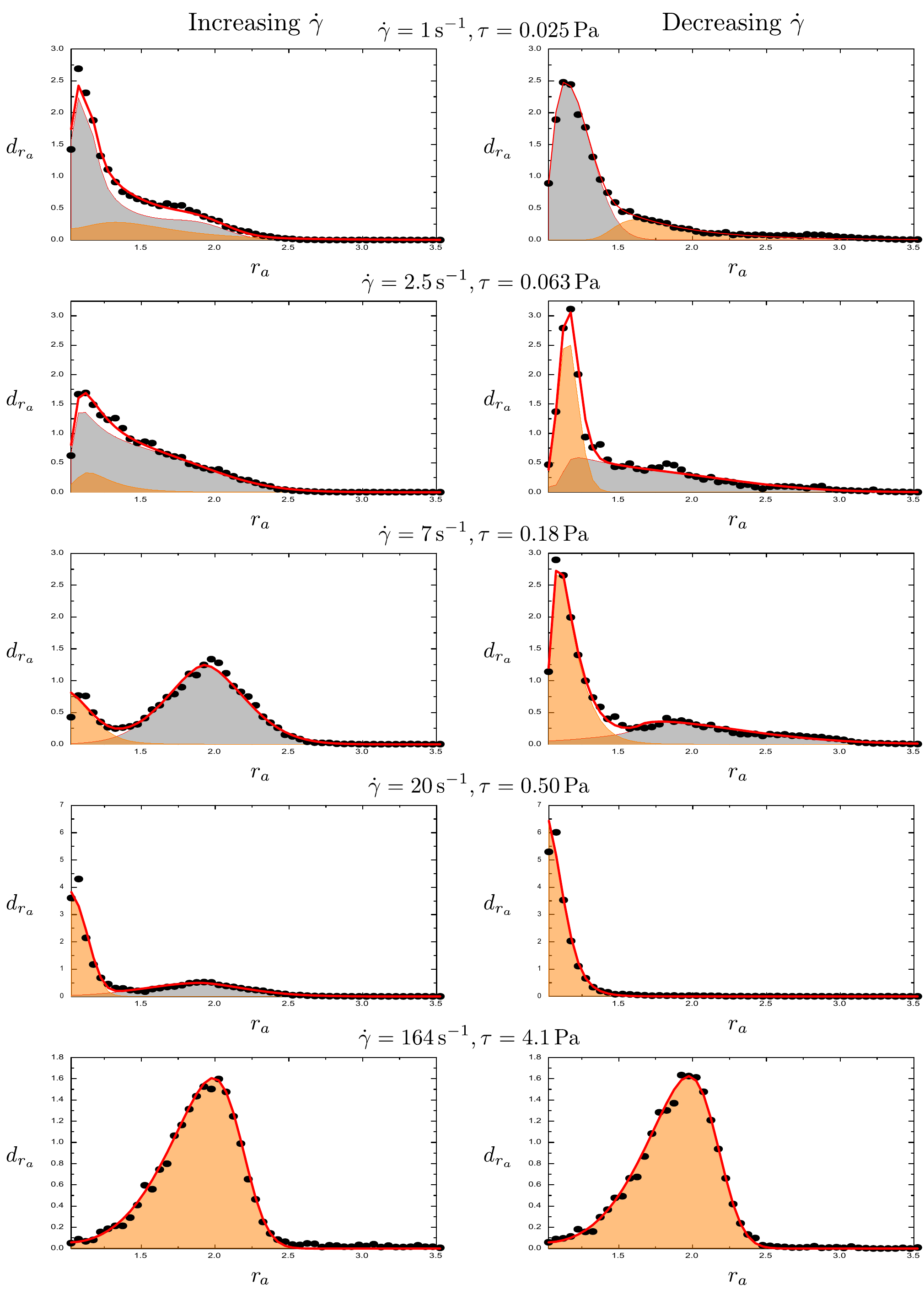}
\caption{Selection of apparent aspect ratio $r_a$ probability distributions for cells in the suspending medium of viscosity 25 mPa.s, for different shear rates. Dots: experimental data; red line: full fit (Eq. \ref{eq:ra}); orange line and area: tank-treading population; red line and grey area: flipping population.}\label{fig:exemples-aspects}
\end{center}
\end{figure}
\subsubsection{Examples of fits}

We first comment on a set of fits obtained in the $\eta_0=25$ mPa.s case for selected shear rates (see Figs. \ref{fig:exemples-angles} and \ref{fig:exemples-aspects}), so as to validate the modelling hypotheses and draw a rough scenario for cell dynamics.  At high shear rates ($\dot{\gamma}=164$ s$^{-1}$ and higher values), it is known that all cells are in tank-treading regime. The apparent aspect ratio distribution is well described by the proposed skew-normal distribution (Fig. \ref{fig:exemples-aspects}, bottom panel), while all angles are close to zero due to strong cell elongation (Fig. \ref{fig:exemples-angles}, bottom panel). The width $w_{\Psi}$ of the angle distribution is of order $1^\circ$ ; as the angle determination is quite precise due to the high apparent aspect ratio, this is an indicator of experimental intrinsic fluctuations: shear chamber vibrations and influence of other cells, which can therefore be considered negligible.

For $\dot{\gamma}=20$ s$^{-1}$, the scenario is different whether $\dot{\gamma}$ is increasing or decreasing. In the latter case, there is still only one population of tank-treading cells, but as the cells are less stretched, their apparent aspect ratio is close to 1. As a consequence, the distribution of their apparent angle has two peaks around 0 and 90$^\circ$, well described by the proposed Cauchy distribution. The two peaks have almost the same height,  indicating an aspect ratio close to 1 (as seen on the $r_a$ distribution), which is the worse case for angle detection. This results in the presence of all angles in the experimental distribution, which would not be well described by a Gaussian distribution, for instance.

On the other hand, in the increasing $\dot{\gamma}$  case,  a second peak appears in the apparent aspect ratio distribution, around $r_a=2$, which is more important for $\dot{\gamma}=12$ s$^{-1}$ or 7 s$^{-1}$, and corresponds to flipping cells with orbit angles close to 0$^\circ$, that is, to rolling. This second peak is well fitted by a  normal distribution, as  hypothesized for  Eq. \ref{eq:raF}. This second peak of rolling cells is present in the decreasing $\dot{\gamma}$ case, but has lower weight. We shall return to this later.

\begin{figure}
\begin{center}
  \includegraphics[width=\columnwidth]{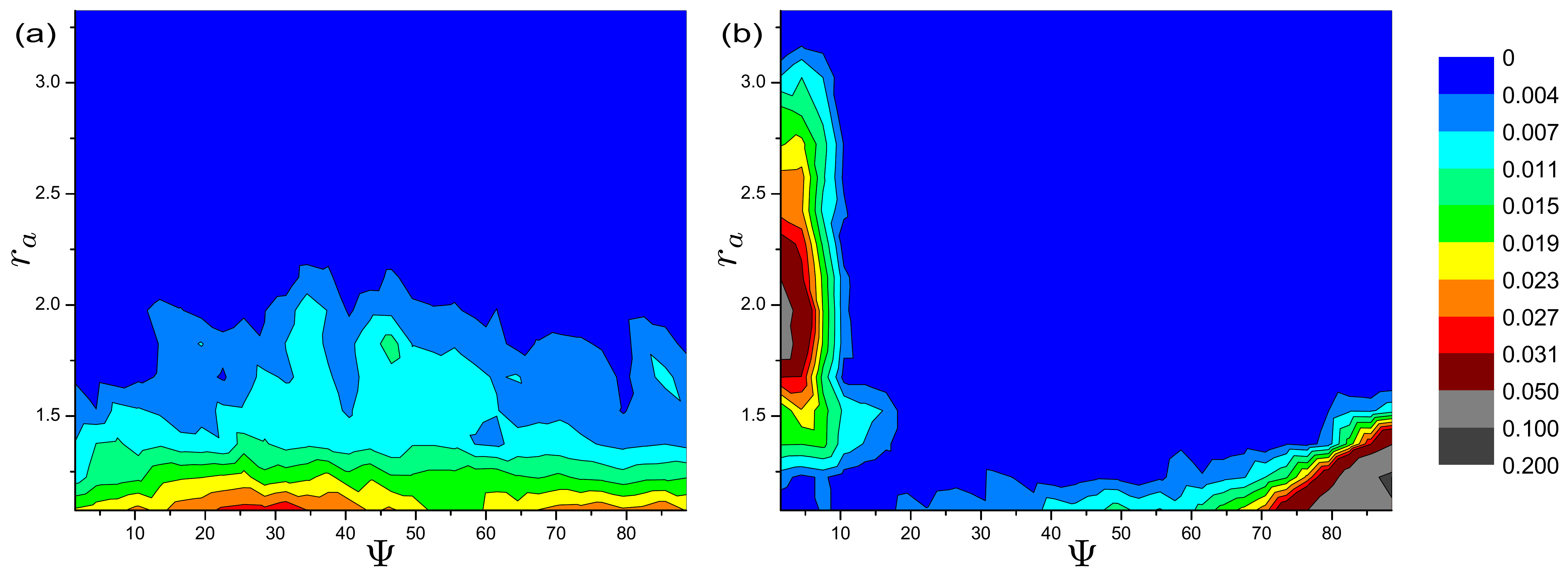}
\caption{Maps of probability densities in the $(\Psi,r_a)$ space. (a): $ \eta_0=25$ mPa.s and $\dot{\gamma}=1$ s$^{-1}$, increasing shear rate case.  (b): $ \eta_0=25$ mPa.s and $\dot{\gamma}=7$ s$^{-1}$, decreasing shear rate case.}\label{fig:dex7low}
\end{center}
\end{figure}

At low shear rates ($\dot{\gamma}$ =2.5 s$^{-1}$ and 1 s$^{-1}$ for this example), the tank-treading cells contribute  in the $r_a$ distribution  with a peak which is now far from 1, with apparent angle is now only around $90^\circ$, along the vorticity direction due to the projection. On the other hand, the contribution of flipping cells is more complex because other orbit angles have appeared. This results in a broadening of angle distribution towards higher angles (from 0 to around $\theta_0^+$), while the aspect ratio distribution  broadens towards $r_a=1$: flipping cells with $\theta_0=0$ are always seen from their edge so the apparent aspect ratio is a peak centered on their 3D aspect ratio $r$, while cells with larger $\theta_0$ also exhibit apparent aspect ratios lower than $r$. Only for $\theta_0=90^\circ$ can this apparent aspect ratio be 1. At the minimum shear rate presented in this example, one can see that the peak of the distribution is not located at 0$^\circ$ anymore. As exemplified by the dashed line in Fig. \ref{fig:rigides}, the location of this peak is indeed an indicator of the value of $\theta_0^-$. For instance, at $\dot{\gamma}$=1 s$^{-1}$, decreasing case, $\theta_0^-$ is around 20$^\circ$ while $\theta_0^+$ is around 50$^\circ$.

In the increasing case particularly, for $\dot{\gamma}$ =1 s$^{-1}$ and 2.5 s$^{-1}$, the angle distribution does not drop to zero for high angles. This cannot be associated only to tank-treading cells with angle 90$^\circ$: a look at the probability distribution in the $(\Psi,r_a)$ space (Fig. \ref{fig:dex7low}(a)) shows that  cells with high angles also have apparent aspect ratios as high as those with low angles, that is, they are also seen edge-on sometimes. This is even clearer when one compares with a case where the high angle cells are in tank-treading regime: in Fig. \ref{fig:dex7low}(b), the aspect ratios of the high angle cells are lower than those of the low angle rolling cells. Note that what matters here is the comparison between the low and high angle populations, not the absolute values of the aspect ratios, that depend on the shear rate through cell deformation.  On the other hand, the angle distribution cannot be described by an equal orbit distribution between $\theta_0^-$ and 90$^\circ$. It requires to divide the population of flipping cells  into two subpopulations, in proportion $p_s$ and $1-p_s$, having orbits between $\theta_0^-$ and $\theta_0^+$, and $\theta_0^+$ and 90$^\circ$, respectively, such that the high angle orbits are less probable. This is exemplified by the dotted line in  Fig. \ref{fig:rigides}, to be compared with the solid line corresponding to equal probabilities. For instance, for $\dot{\gamma}=1$ s$^{-1}$, increasing shear rate case, we find $p_s=0.81$ such that the ratio $\alpha_{u/s}$ between the probability of an orbit  with $\theta_0^+<\theta_0<90^\circ$, and of that of an orbit with $\theta_0^-<\theta_0<\theta_0^+$, is equal  here to 0.22.  $\alpha_{u/s}$ is defined by $[(1-p_s)/(90-\theta_0^+)] / [p_s/(\theta_0^+-\theta_0^-)]$.

\begin{figure}
\begin{center}
  \includegraphics[width=\columnwidth]{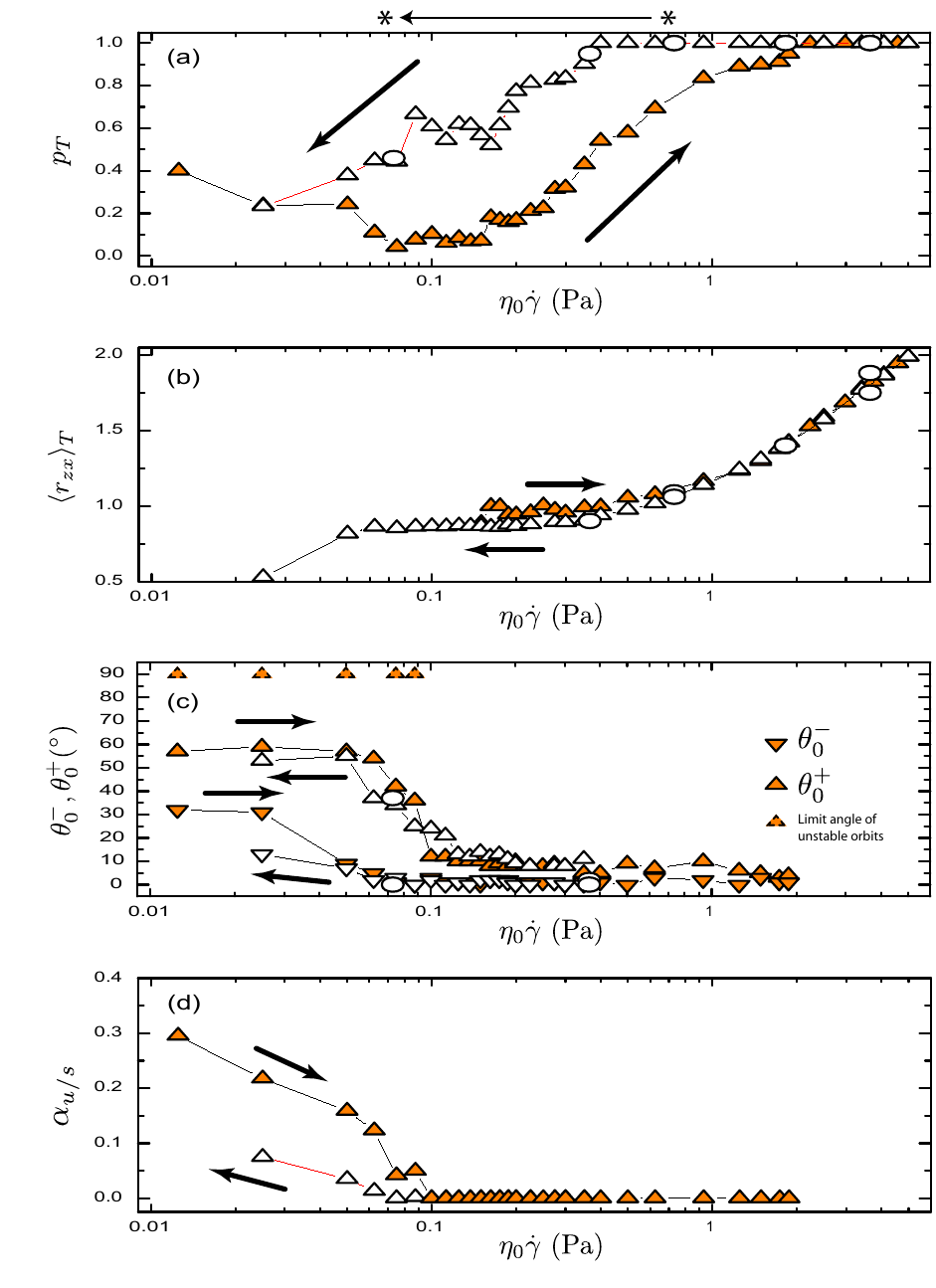}
\caption{Main parameters characterizing cell dynamics for $\eta_0=25$ mPa.s as a function of shear stress $\tau=\eta_0 \dot{\gamma}$  in both increasing $\dot{\gamma}$ (full triangles) and decreasing $\dot{\gamma}$  (empty triangles)  cases, as indicated by arrows. Empty circles: additional control experiments  at decreasing $\dot{\gamma}$ made with RBCs from two other donors and a suspending medium prepared independently. On top of panel (a), the transition between the two stresses indicated by stars corresponds to that studied in Fig. \ref{fig:intermittent}. (a) Proportion $p_T$ of cells in tank-treading like motion.  (b) Mean apparent aspect ratio $\langle r_{zx}\rangle_T=r_T + w_T \,\alpha \, \sqrt{2}/\sqrt{\pi(1+\alpha^2)}$ of tank-treading cells.  (c) Minimal and maximal orbit angles  $\theta_0^-$ and $\theta_0^+$ for stable flipping cells. Dotted symbols indicate the limit orbits (90$^\circ$) of (probably) unstable orbits between $\theta_0^+$ and $90^\circ$, when present. (d) Ratio $\alpha_{u/s}$ between the probability of  orbits  with orbit angle between $\theta_0^+$ and $90^\circ$, and between $\theta_0^-$ and $ \theta_0^+$.}\label{fig:data2+7}
\end{center}
\end{figure}

\subsubsection{Transition scenario}

In Fig.~\ref{fig:data2+7} Êwe show the evolution with  flow stress $\tau=\eta_0\dot{\gamma}$ of the main parameters characterizing the  collective dynamics of red blood cells. 

The main difference between the increasing and the decreasing $\dot{\gamma}$ cases lies in the  proportion $p_{T}$ of cells in tank-treading regime (Fig.~\ref{fig:data2+7} (a)).  A full hysteresis loop is highlighted: at low and high flow stresses, the amount of cells in tank-treading regime are identical. At low flow stress, we observe that, according to our model, the proportion of tank-treading cells is not equal to 0.  We shall propose later an interpretation of this result, which calls for a refinement of the model.

For increasing $\dot{\gamma}$, a significant increase in the proportion of tank-treading cells is observed  from  $ \tau_{c-\min}^>=0.08$ Pa until all cells are in tank-treading regimes at $\tau_{c-\max}^>=2$ Pa. This shows that  the dispersion in the transition stress towards tank-treading $\tau_c^>$ is high: the transition values span over one decade. This result is in agreement with \cite{fischer13}, where for $\eta_0=24$ mPa.s a transition to tank-treading is found around $\tau=0.23$ Pa but with a smooth transition that indeed spans over one decade, as here (see Fig. 2(b) in \cite{fischer13}).

For decreasing $\dot{\gamma}$, the first cells to leave the tank-treading regime do it at a flow stress $\tau_{c-\max}^<$=0.4 Pa smaller than   $\tau_{c-\max}^>$. The last cells to leave the tank-treading regime seem to do it around $\tau_{c-\min}^<$=0.02 Pa. The values for $\tau_{c}^>$ and $\tau_{c}^<$  found in \cite{dupire12} and \cite{mauer18} lie in the lower part of the transition zone we find.

Fig. \ref{fig:data2+7}(b) shows the mean aspect ratio of cells in tank-treading regime, considering the aspect ratio $r_{zx}$ between the axis along the flow direction and the axis in the vorticity direction, whose distribution is given by Eq. \ref{eq:skew}. This mean value is a function of the fit parameters through $\langle r_{zx}\rangle_T=r_T + w_T \,\alpha \, \sqrt{2}/\sqrt{\pi(1+\alpha^2)}$ and is directly linked to the deformability of cells. Due to the dispersion in cell properties (and, probably, to swinging motion), the standard deviation around this mean value is between 0.1 and 0.15 within the whole stress range.

Deformation is quite similar for both directions of shear rate variation.  However, in the hysteretic region for the tank-treading population, the apparent aspect ratio is up to 10\% smaller in the decreasing $\dot{\gamma}$ case and is almost constant while it varies more strongly for higher stress values. On the other hand, in this range stress is not so small that we should expect cell shape not to be modified by the flow. Indeed, even for lower stresses transitions between different regimes occur and orbit angles of the flipping regimes change  continuously: another explanation for this plateau has to be found.

We note that, while the aspect ratio of cells in tank-treading regime should be an increasing function of stress, the population of cells in that regime is not constant. We hypothesize that the plateauing shows that, upon an increase of shear rate, the contribution of the cells switching to tank-treading compensates the increased deformation of the cells already in tank-treading regime. This implies that the latter  are more deformable (in the sense of stretchable) than the newcomers, which are themselves the most deformable of the flipping cells that have not switched  to tank-treading yet. This is coherent with the  picture of more deformable cells needing smaller stress to make their transition.
Similar reasoning shows that, upon a decrease of shear rate, the less deformable cells of the tank treading population are the first ones to transit towards the flipping motion. 

\begin{figure}
\begin{center}
  \includegraphics[width=\columnwidth]{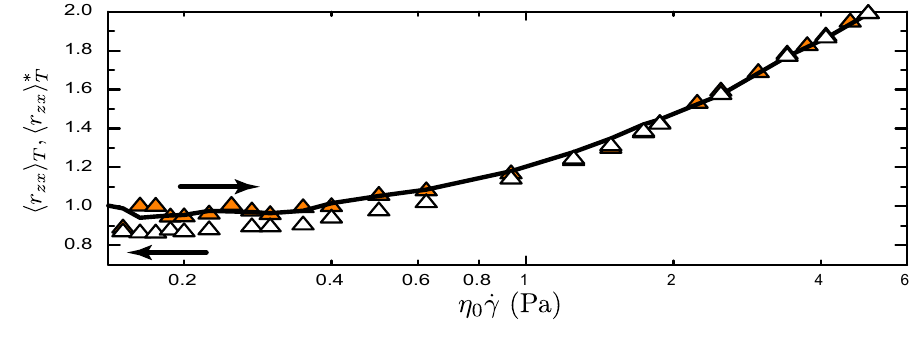}
\caption{Mean apparent aspect ratio $\langle r_{zx}\rangle_T=r_T + w_T \,\alpha \, \sqrt{2}/\sqrt{\pi(1+\alpha^2)}$ of tank-treading cells, in the increasing and decreasing shear case (same data as in Fig.\ref{fig:data2+7}(b)). Additional full line shows the mean apparent aspect ratio $\langle r_{zx}\rangle_T^*$ when the less deformable cells are removed from the distributions in the decreasing shear rate case (see text).}\label{fig:renorm}
\end{center}
\end{figure}

A consequence of this would be that at a given stress, because of the hysteresis in the transition thresholds, the tank-treading population in the decreasing $\dot{\gamma}$ case contains more cells than in the increasing $\dot{\gamma}$ case. To validate the whole picture, let us first denote by $P_{T\nnearrow}$ and $P_{T\ssearrow}$ the proportion of cells in the tank-treading regime in the increasing and decreasing $\dot{\gamma}$ cases, respectively. From the previous discussion, we conclude that, among the tank-treading cells in the decreasing $\dot{\gamma}$ case, the more deformable ones are also in tank-treading regime in the increasing shear rate case, while the less deformable are not. The proportions of these two subpopulations are, by definition, $P_{T\nnearrow}/P_{T\ssearrow}$ and $1-P_{T\nnearrow}/P_{T\ssearrow}$, respectively. Considering, for a given stress $ \tau$, the whole skew-normal distribution of $r_{zx}$ in the decreasing shear rate case, we make a cut-off in this distribution by considering only the last $P_{T\nnearrow}/P_{T\ssearrow}$ cells, on which the mean value of  $r_{zx}$, that we denote $\langle r_{zx}\rangle_T^*$, is calculated. The result is shown in Fig. \ref{fig:renorm}. The curve of the mean value taken on the more deformable cells in the decreasing shear rate case now collapses onto the mean value $\langle r_{zx}\rangle_T$ in the increasing shear rate case, thus validating our hypothesis. 


Even in the hysteresis domain, the main characteristics of the cell flipping dynamics remains unchanged whether $\dot{\gamma}$  is increased or decreased (Fig.~\ref{fig:data2+7} (c)): the range of orbit angles of flipping cells does not depend on shear rate history. This is in agreement with \cite{dupire12}, where it is found that once the cells have left the tank-treading regime when decreasing $\dot{\gamma}$, they follow the same dynamics as the one they had at the same shear rate in the increasing sequence. However, in \cite{dupire12}, no quantitative data on the orbit angles upon a decrease of $\dot{\gamma}$  are given. A single example is given (Fig 4C), where an orbit angle around 30$^\circ$ is shown. Here, we do not prove formally that a given cell will follow the same orbit for a given stress, whatever the shear rate history, but the similarities in the angle distribution for both shear rate variation cases pleads in favor of such a scenario. Finally, the large range of orbits is another signature of the dispersion in cell properties.

Fig.~\ref{fig:data2+7} (d) shows that at low stress the flipping populations must be divided into two subpopulations with different probabilities. Interestingly, the range of existence of a population reaching orbits up to 90$^\circ$ (in probability $1-p_s$) is similar to that where $\theta_0^+$ is almost constant (between 50$^\circ$ and 60$^\circ$, when $\tau \lesssim$ 0.07 Pa close to the upper bound for $\eta_0\dot{\gamma}=0.05$ Pa found in \cite{dupire12}). This shows that, while limit orbit angles $\theta_0^-$ and $\theta_0^+$ vary smoothly with stress, a population with many different orbits equally distributed between $\theta_0^+$ and $90^\circ$, with $\theta_0^+$ very different from $90^\circ$, emerges (or disappears) as a whole as the stress is varied. This is in strong agreement with the observation in \cite{dupire12} of the existence of a threshold in $\theta_0$ above which orbits are not stable and are rather distributed on a wide range (hence a much lower relative probability $\alpha_{u/s}$).  Here we exhibit, with strong statistical weight, that this threshold is $56^\circ \pm 2^\circ$ (averaged on values for $\tau\le 0.05$ Pa). Note finally that at decreasing $\dot{\gamma}$, cells also reach this state of unstable Jeffery orbits, which was not mentioned in \cite{dupire12}. There is a  difference in the number of cells in unstable orbits depending on the direction $\dot{\gamma}$ variation, which is even clearer in the case of low $\eta_0$ discussed in the next section. 

According to the fitting procedure, we find a non negligible amount of tank-treading cells at the lowest stresses, which clearly does not  correspond to the reality but must be seen as an artifact of the chosen model. As seen in Fig. \ref{fig:exemples-angles}, these cells are seen with apparent angle centered on $90^\circ$. This is consistent with the assumption that the unstable flipping cells do not fully follow Jeffery orbits but spend more time aligned with the flow, presenting thus a kind of transient tank-treading regime, as found in physiological solution in \cite{goldsmith72}. This interpretation is reinforced by the overlap of the domain of existence of these tank-treading-like cells and of unstable orbits ($\tau\lesssim 0.08$ Pa in the increasing $\dot{\gamma}$ case). This feature as a strong impact on the overall distribution of aspect ratios and apparent angles. \\

While the decrease of the flipping cell population with increasing $\dot{\gamma}$ is associated with a decrease of the orbit angle $\theta_0$, the data do not clearly show in which orbits the cells are right before switching to tank-treading. In \cite{dupire12}, the tracking of single cells showed that all studied cells are in rolling regime before they do so. To connect this observation to our statistically relevant sample, we compare our data for the population $1-p_{T}$ of cells in flipping regime to that expected from the following model: we assume that, for a given relative stress variation $d\tau/\tau=d \log(\tau)$, the flipping cells having orbit angle comprised between 0 and $d\theta_t$, if any, switch towards tank-treading. The ratio  $q=d\theta_t/d \log(\tau)$ is considered as a constant to be determined. The theoretical proportion of cells in flipping  regime $1-p_{T,th}$ then obeys \begin{equation}
1-p_{T,th}(\tau+d\tau)=\big[1-p_{T,th}(\tau)\big]\times \big[1-p_s(\tau)\frac{\max\Big(0,\min\big(\theta_0^+(\tau),q \,d \log (\tau)\big)-\theta_0^-(\tau)\Big)}{\theta_0^+(\tau)-\theta_0^-(\tau)}\big].\end{equation}

\begin{figure}
\begin{center}
  \includegraphics[width=\columnwidth]{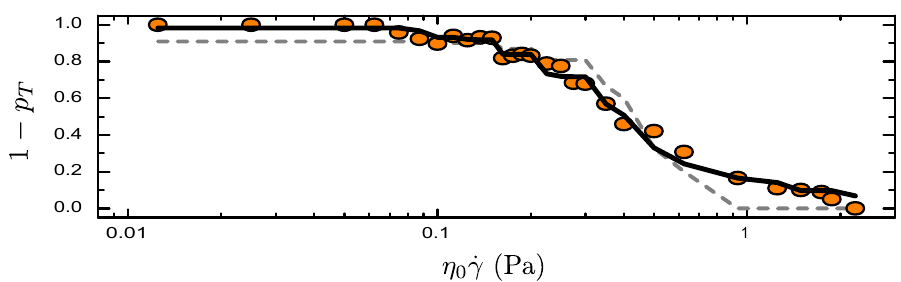}
\caption{Evolution of the proportion $1-p_{T}$ of cells in flipping regime  for $\eta_0=25$ mPa.s as a function of shear stress $\tau=\eta_0 \dot{\gamma}$, for the increasing shear rate case; dots: experimental data (same as  \ref{fig:intermittent}(a) with the low stress values set to 1 to discard the fake tank-treading cells); full line: fit with equation \ref{eq:transitTT}. Dashed line: best fit with a constant transition angle by  absolute stress variations $d \theta _t/d\tau$ (see text).}\label{fig:transitTT}
\end{center}
\end{figure}

If this equation is discretized on the experimental stresses $\tau_i$, $i\le n$, one finds

 \begin{multline}1-p_{T,th}(\tau_{i+1})=\big[1-p_{T,th}(\tau_1)\big]\\
 \times \prod_{j=2}^n\big[1-p_s(\tau_{i-1})\frac{\max\Big(0,\min\big(\theta_0^+(\tau_{j-1}),q \,\log (\tau_{j}/\tau_{j-1})\big)-\theta_0^-(\tau_{j-1})\Big)}{\theta_0^+(\tau_{j-1})-\theta_0^-(\tau_{j-1})}\big],\label{eq:transitTT}\end{multline}

where $1-p_{T,th}(\tau_1)$ is the proportion of cells in flipping regime at the lowest explored stress, expected to be equal to 1 in our case where we have explored low enough stresses. Together with $q$, they constitute the free parameters of the model, that can be adjusted to fit the experimental data. As shown in Fig. \ref{fig:transitTT}, a good fit is obtained, with $q= 10.7 ^\circ$. This means that, upon a relative stress variation $d\tau/\tau$ of 5\%, only the cells with orbit angle between 0 and $0.54^\circ$ will switch to tank-treading. This validates the observation made in \cite{dupire12}. Note that a model that would assume a constant transition angle by  absolute stress variations $d \theta _t/d\tau$ does not result in a good fit with the experimental data (see Fig. \ref{fig:transitTT}).

For  decreasing $\dot{\gamma}$, Fig.~\ref{fig:data2+7} (c) directly shows that at least half of the cells switching from tank-treading to flipping do switch to an orbit close to rolling ($\theta_0^+ <10^\circ$, in the range $0.1$ Pa$<\tau<0.4$ Pa), while at lower stress the others may  a priori directly reach other orbits. However, the same analysis made with equation \ref{eq:transitTT} on the data for  decreasing $\dot{\gamma}$ reveals that $q=16.6 ^\circ$ that is, 
they
always reach an orbit close to rolling. \\

\begin{figure}
\begin{center}
  \includegraphics[width=\columnwidth]{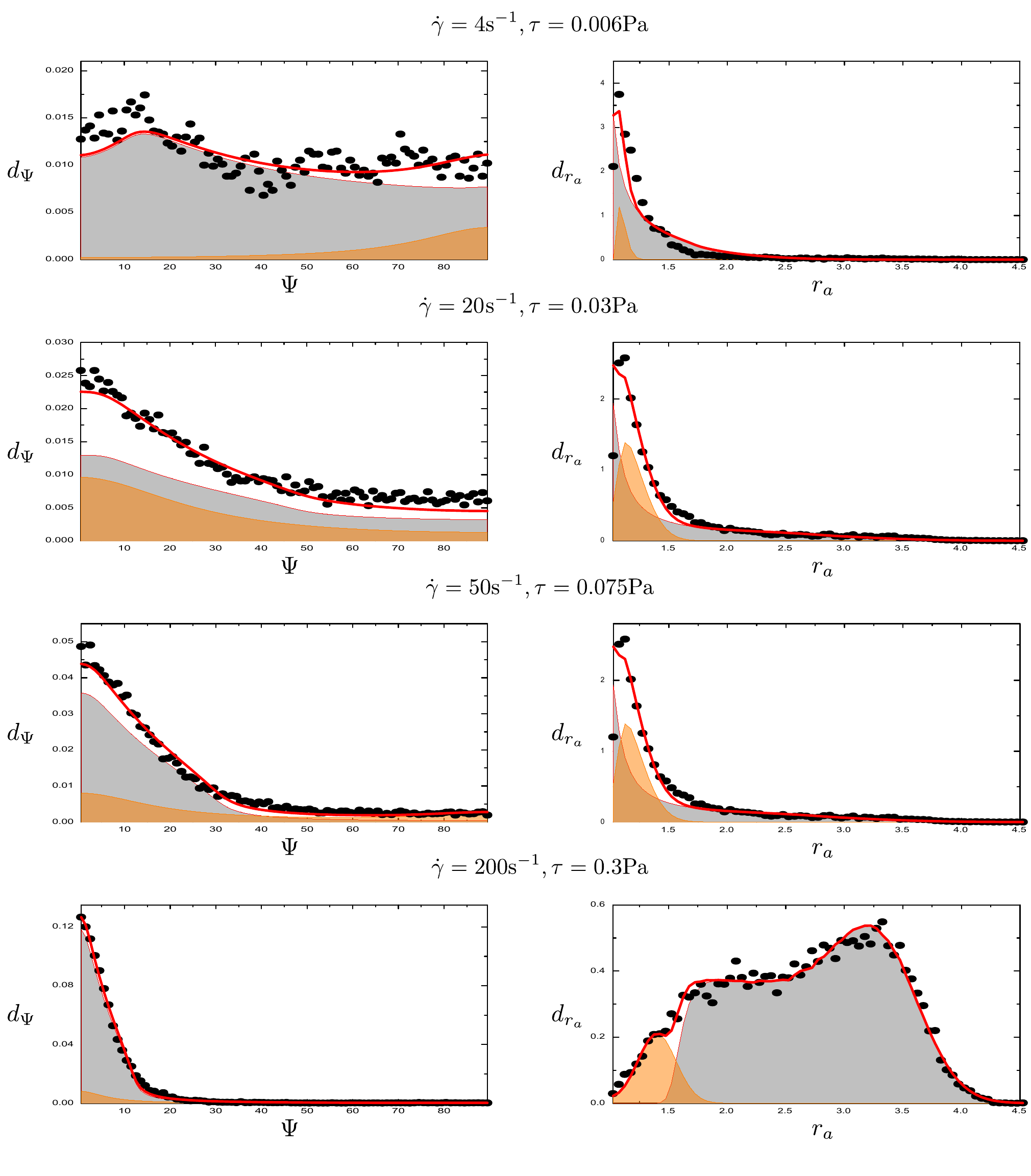}
\caption{Selection of apparent angle $\Psi$ and apparent aspect ratio $r_a$ probability distributions for cells in the suspending medium of viscosity 1.5 mPa.s, for different increasing shear rates. Dots: experimental data; red line: full fit (Eqs. \ref{eq:ra} and \ref{eq:Psi}); orange line and area: tank-treading population; red line and grey area: flipping population.}\label{fig:exemples-dex0}
\end{center}
\end{figure}

\subsection{Cells in a fluid of physiological viscosity}

\subsubsection{Examples of fits}

As for the first case, we start by commenting on  some selected distributions and the corresponding fits, shown in Fig. \ref{fig:exemples-dex0} (increasing shear rate case). At the highest explored shear rate $\dot{\gamma}=200$ s$^{-1}$, the apparent angles mainly span between 0 and 10$^\circ$, indicating flipping orbit angles between 0 and 10$^\circ$. For such orbits, the apparent aspect ratio spans between a value larger than 1 and $r$ the cell aspect ratio (grey envelope in Fig. \ref{fig:exemples-dex0}, bottom panel). Here, we observe that the probability of apparent aspect ratio  around 1 is not 0, indicating that tank-treading like motion is present, as found by the fitting procedure.

For lower $\dot{\gamma}$, the orbit distribution is wider, but a contribution of tank-treading like motions is still needed to explain the whole distribution. At shear rate $\dot{\gamma}=20$ s$^{-1}$, all orbit angles are present, but the contribution of high $\theta_0$ orbits is less important, 
indicating probably the presence of unstable high angle orbits (see dotted curve of Fig. \ref{fig:rigides}, to be compared with the solid curve on the same figure). At even smaller shear rate  $\dot{\gamma}=4$ s$^{-1}$, these angles now have similar probabilities, but small angle orbits are not present anymore, as on the dashed curve of Fig. \ref{fig:rigides}.

\begin{figure}
\begin{center}
  \includegraphics[width=\columnwidth]{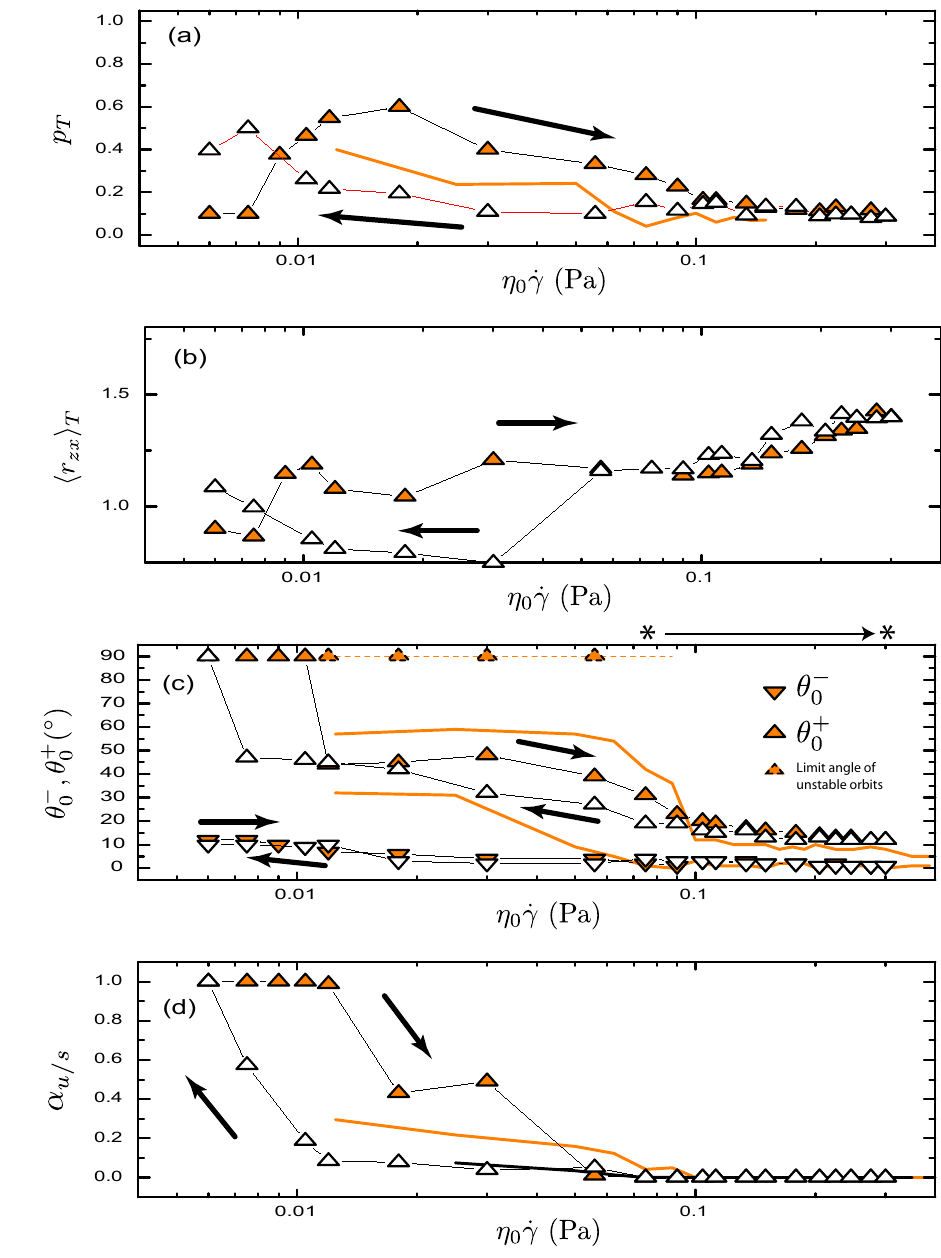}
\caption{Main parameters characterizing cell dynamics for $\eta_0=1.5$ mPa.s as a function of shear stress $\tau=\eta_0 \dot{\gamma}$  in both increasing (full triangles) and decreasing (empty triangles) $\dot{\gamma}$ cases, also indicated by arrows.  (a) Proportion $p_T$ of cells in tank-treading like motion. For sake of comparison, the full line corresponds to the data for $\eta_0=25$ mPa.s, increasing $\dot{\gamma}$ case, already shown in Fig. \ref{fig:data2+7}(a). (b) Mean apparent aspect ratio $\langle r_{zx}\rangle_T$ of tank-treading cells. (c) Minimal and maximal orbit angles  $\theta_0^-$ and $\theta_0^+$ for flipping cells. Dotted symbols indicate the limit orbits of (probably) unstable orbits between $\theta_0^+$ and $90^\circ$. The lines correspond to the data for $\eta_0=25$ mPa.s, increasing $\dot{\gamma}$ case, already shown in Fig. \ref{fig:data2+7}(c). On top the panel, the transition between the two stresses indicated by stars corresponds to that studied in Fig. \ref{fig:transit0detail}. (d) Ratio $\alpha_{u/s}$ of the probability of an orbit  with orbit angle between $\theta_0^+$ and $90^\circ$, and of that of an orbit between $\theta_0^-$ and $ \theta_0^+$. The full lines correspond to the   data for $\eta_0=25$ mPa.s, increasing (orange) and decreasing (black) $\dot{\gamma}$ cases.}\label{fig:data0}
\end{center}
\end{figure}

\subsubsection{Transition scenario}

Fig. \ref{fig:data0} details the values of the main parameters while the shear rate is varied. As for the high $\eta_0$ case, the range of $\theta_0$ increases as the shear rate becomes lower, and it does not depend on the direction of $\dot{\gamma}$ variation (Fig. \ref{fig:data0}(c)). Interestingly, the threshold in flow stress below which orbits other than those close to rolling (maximal orbit angle $\theta_0^+$  larger than  10$^\circ$) is similar to that obtained for a much more viscous suspending fluid.  For smaller stress, a plateau in  $\theta_0^+$  is also reached, at  45$^\circ\pm 2^\circ$ while orbits between $\theta_0^+$ and $90^\circ$ with smaller relative probability $\alpha_{u/s}$ appear (Fig. \ref{fig:data0}(d)). We interpret these orbits as unstable orbits. For small stresses, all orbits between $\theta_0^-$ and $90^\circ$ become equiprobable ($\alpha_{u/s}=1$). In that case $\theta_0^+$ is ill-defined (any value can yield the same fitting function eventually, with an adapted value of $p_s$). If we interpret this equiprobability as a signature of the stability of the large angle orbits, then $\theta_0^+=90^\circ$, which is the choice made in Fig. \ref{fig:data0}(c). 

The range of existence of unstable orbits is not the same whether we consider increasing or decreasing $\dot{\gamma}$ Fig. \ref{fig:data0}(d). This feature, which has not been observed in the literature so far, is also a key ingredient to understand the variations in the proportion of tank-treading like cells  (Fig. \ref{fig:data0}(a)) and the associated mean aspect ratio (Fig. \ref{fig:data0}(b)).

The hysteretic behaviour is seen for stresses below $\sim 0.05$ Pa. Let us comment first on  what happens above this threshold.  There, the mean aspect ratio of tank-treading like cells is much larger than 1 (Fig. \ref{fig:data0}(b)), which would support the observation made in \cite{lanotte16} and \cite{mauer18}: in physiological conditions, the authors  observe that between  40  s$^{-1}$ and 200  s$^{-1}$, more than 60\% of cells are rolling or vacillating-breathing stomatocytes. This feature disappears for a shear rate of 40  s$^{-1}$, corresponding in their case to a stress of 0.04 Pa. This is also the value around which we observe a strong drop of the mean aspect ratio, for decreasing $\dot{\gamma}$. This drop is not observed for increasing $\dot{\gamma}$ because of overlapping with another phenomenon, which we discuss now: when increasing $\dot{\gamma}$, unstable orbits are present in the range $0.01$ Pa $ <\tau< 0.05$ Pa (Fig. \ref{fig:data0}(c)), which is  very close to that where the proportion of (apparently) tank-treading like cells is non negligible (Fig. \ref{fig:data0}(a)), and associated apparent aspect ratio still large (Fig. \ref{fig:data0}(b)). This observation holds  also when decreasing $\dot{\gamma}$ for the domain $\tau<$ 0.01 Pa.

The similarity in the range of existence of tank-treading like cells  and unstable orbits, already observed in the high $\eta_0$ case, indicates that unstable (large angle) orbits do not follow completely Jeffery orbits but longer time is spent aligned with the flow, as observed in one of the few experiments made in that regime (\cite{goldsmith72}).

The threshold  $\tau \sim 0.01$ Pa for the appearance of the  first feature in the increasing $\dot{\gamma}$ case (considered also in   \cite{goldsmith72}) is the same here as in this historical paper, and we also observe around this threshold the apparition of rolling motion (while below this threshold $\theta_0^-$ is slightly larger than 10$^\circ$), as well as the loss of stability of orbits close to tumbling (high orbit angles).

\subsection{The whole picture}

We have synthesized in Fig. \ref{fig:basicdiagram} the main findings of this study in terms of transition threshold and existence intervals. They are compatible with those mentioned in the main experimental studies of the literature as commented along the above analysis. The sole notable incompatibility is the existence of off-plane motion in flipping regime in the case of high $\eta_0$, which was not observed in \cite{levant16} while observed in \cite{dupire12}, but the parameter space is different.

Many features are quantitatively similar in the low and high viscosity cases, as soon as one considers the flow stress $\tau=\eta_0 \dot{\gamma}$ as the control parameter (see full lines in Fig. \ref{fig:data0}(a,c,d)). This observation was also made in \cite{mauer18}, where stresses  larger than here were considered and other regimes and deformation observed. 

The range of existence of flipping orbits, and  the values of the angles of stable orbits,  are a remarkably robust feature that does not depend on the history of shear rates and only weakly on the external viscosity for a considered stress (the angles are shifted by about 10-15$^\circ$ in the high viscosity case). Flipping motions are well described by Jeffery orbits, which seem stable as long as  the orbit angle is lower than $\theta_0^+\sim 50^\circ$. There exists a range where higher angles orbits appear but seem to be unstable: all angles between $\theta_0^+$ and 90$^\circ$ become available, with small probability. The existence of many orbits is in contradiction with recent numerical simulations \cite[see][]{cordasco13,cordasco14_2,sinha15}, where it is often found, for small or large values of $\eta_0$, that stable orbits are those with angle $\theta_0$ close to 0$^\circ$ or $90^\circ$, depending on the paper considered (see  Supplemental Material for  a detailed description). However, more recent results by \cite{mendez18} exhibit stability of other orbits with a drift towards rolling as the shear rate is increased.

Unstable orbits are present below a threshold $\sim 0.04$ Pa for both viscosities, in higher proportion at low viscosity, and associated with strong hysteresis with stress variation (through $\dot{\gamma}$). On average, the unstable-to-stable transition occurs for larger stress than the stable-to-unstable transition. It is particularly clear at low $\eta_0$, while at high $\eta_0$ these transitions are close to the lowest explored value, making the picture less obvious.  Unstable orbits are modified Jeffery orbits: longer time is spent aligned with the flow, which is seen through an artificial increase in the number of tank-treading cells in our model. At low stresses (only explored in the low viscosity case, though), all orbits become equiprobable, as for rigid cells, though low angles are not explored any more. Indeed, normal cells never behave like rigid cells even under small stress.

Finally, in the tank-treading regime (at high $\eta_0$), cell deformation is similar whatever the history of shear rates (Fig. \ref{fig:renorm}).  At low $\eta_0$ and high stress, a signature of the presence of stomatocytes is detected. This feature is also history-independent.

Contrary to the previous features, the flipping to tank-treading transition strongly dependens on $\eta_0$. While tank-treading appears for  $\tau$ between 0.1 and 2 Pa at high $\eta_0$ (with the more stretchable cells transiting first), the threshold is expected to be at least 2 orders of magnitude higher at low $\eta_0$ \cite[see][]{morris79}. Note however that in that case, according to our results and \cite{lanotte16}, the notion of such a transition would not be as well defined because of the strong deformations undergone by the cells, whose shapes strongly depart from the usual discocyte.

To summarize, the characteristics of each regime (accessible orbits and stability  of orbits for the flipping regime, deformation in the tank-treading regime) are controlled by the flow stress with no dependency with the shear rate history, while the possibility of transition towards tank-treading also depends on the viscosity ratio. An hysteresis is associated with this latter transition, as well as with the unstable-to-stable Jeffery orbit transition.\\

\subsection{Dispersion in the transition thresholds}

The width of the transition zones between the different regimes or between the different orbits can be associated with the dispersion in RBC properties. Cells can, in particular, differ by their size, deflation (volume/surface ratio), and mechanical properties: viscosity of the haemoglobin solution, shear and bending moduli of the membrane, stress-free shape or spontaneous curvature. Linking the width of the transition zone to a dispersion in all these parameters would require a full (numerical or theoretical) model that is not available for now. Yet, we can draw some conclusion from our results and from the partial results found in the literature.

We first focus on the rolling to tank treading transition for high $\eta_0$. There is a factor $\sim 20$ between the  two stresses between which the transition of all cells occur (in both increasing and decreasing $\dot{\gamma}$  cases).

The effect of bending rigidity on the transition towards tank-treading has been explored in \cite{yazdani11} where the bending modulus was varied by a factor 25: the critical stress for transition varies only by a factor 1.7. In \cite{mendez18}, it is argued that because of the strong dependency of the transition thresholds on the choice of stress free shape (relatively to the value of in-plane shear elasticity), out-of-plane deformations (that would be controlled by the bending modulus)  should play a minor role. This is confirmed by the  reproduction of several important features (orbital drift and transition towards tank-treading) through a shape preserving model. Finally, for vesicles where shear modulus vanishes, it  as also been shown in \cite{farutin12_2} that the transition towards tank-treading depends only weakly on the bending modulus. We conclude that dispersion in bending moduli, that seem to lie between $3\times10^{-19}$ and $9\times10^{-19}$ N.m \cite[see][]{betz09,evans08,sinha15}, is unlikely to contribute significantly to the dispersion in transition stresses.

As already mentioned, the viscosity of the cytoplasm can lie between 6 and 20 mPa.s within one sample. From the results of \cite{fischer13} shown in Fig. \ref{fig:basicdiagram}, the critical stress varies by one decade when the external viscosity (and therefore, the ratio between internal and external viscosities) is varied by one decade. Therefore, variations in inner fluid viscosity may account for a factor around 3 in the critical stress dispersion.

The flow stress can be directly compared to the shear stress in the membrane, that is proportional to $\mu/R$, where $\mu$ is the shear modulus and $R$ a typical size of the cell. In \cite{henon99}, it is stated that according to the literature $\mu$ lies between 4 and 10 \textmu N/m ; they themselves find values between 1 and 4 \textmu N/m  (but extreme values are scarcer) while in \cite{mills04} values between 5 and 11 \textmu N/m are found. Beyond the question of the exact value of this modulus, that indeed depends on the chosen model for elasticity, one can see that it can be dispersed by a factor 2 to 3, which can be multiplied by a factor 1.3 if one takes into account that cell size can vary by this amount: cell diameters are typically between 7 and 9 \textmu m (\cite{canham68}).

In \cite{peng14}, the effect of stress-free shape on the transition threshold is explored, and a factor 2  in this threshold arises when the this shape is varied within an admitted range. In \cite{sinha15}, the effect of spontaneous curvature on the transition  have been explored, but only two cases of interest are explored, with very tiny variation in the transition threshold. Finally, in \cite{cordasco13}, oblate ellipsoids with mechanical properties similar to that of RBCS are considered and is is shown that within a range of aspect ratio between 1.2 and 2, the transition threshold cannot vary by more than 2, but the high discretization in the explored parameter space prevents from drawing precise conclusions. In addition, the reduced volume (volume divided by the volume of a sphere of same surface) of a red blood cell is typically between 0.61 and 0.76 \cite[see][]{canham68}, which corresponds for an oblate ellipsoid to $r$ between 2.7 and 3.7, far from the explored range. 

All these considerations seem to indicate that taking into account all variabilities in cell properties is necessary to account for the large dispersion in transition threshold. This also implies that these variations are necessarily independent if one wants to account for the factor 20 that characterizes that dispersion.\\

Other transitions are associated with variations within the flipping regime: unstable to stable orbits, flipping to rolling, etc.  We generally observe a dispersion in the transition stress characterized by a factor of order 3 to 10. This tends to prove that not all cell parameters are involved in this transition. It is suggested in \cite{dupire12} than in-plane elasticity (and not dissipation) is responsible for the orbital drift, and in \cite{mendez18}, it is shown, as in our study, that the flipping dynamics depends very weakly on the fluids viscosities. This removes variations in the viscosity of the haemoglobin solution from the cause of dispersion, and may explain why the dispersion in the transitions within the flipping regime is smaller than for the rolling to tank treading transition. 

\begin{figure}
\begin{center}
  \includegraphics[width=\columnwidth]{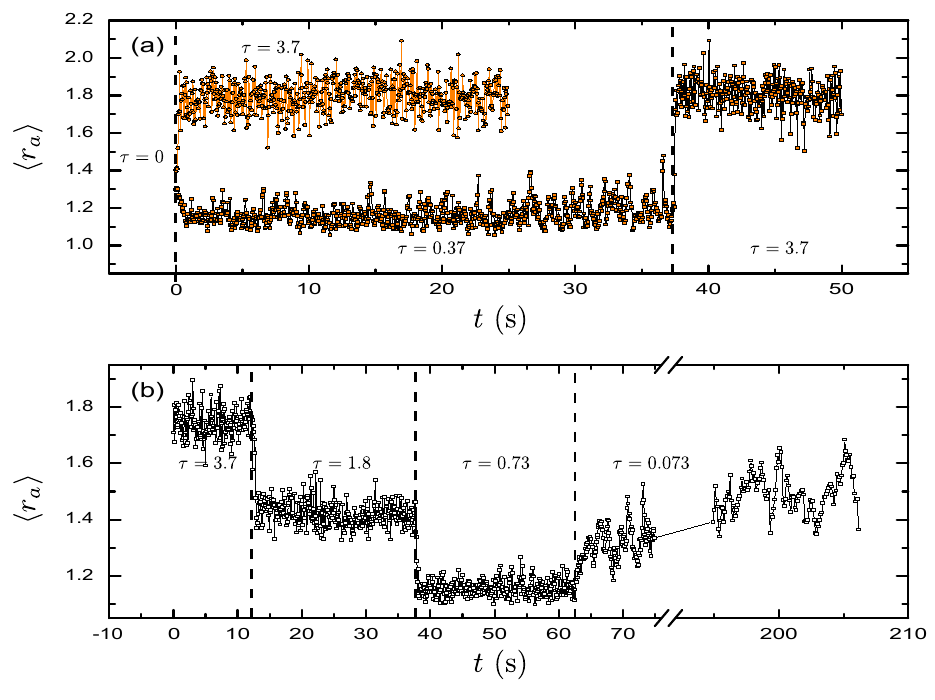}
\caption{Examples of transition dynamics for parameter $\langle r_a\rangle$ after shear stress sharp increases (a) or decreases (b). $\eta_0=25$ mPa.s. Shear stresses $\tau$ are in Pa. Vertical dotted lines correspond to the shear stress instantaneous switches (by changing the shear rate). (a) Two-step or one-step increase from 0 to 3.7 Pa. (b) Three-step decrease from 3.7 Pa to 0.073 Pa.}\label{fig:trans2+7}
\end{center}
\end{figure}

\subsection{Transition times and intermittent regimes}
\label{sec:transit}

Transition times between states are a priori a function of the initial stress, the final stress, but also on the state in which the cells are, which may depend on the history of shear rates. While full study of these times is out of the scope of the present work, a few robust points deserve to be mentioned.

In experiments presented above, $\dot{\gamma}$  was varied by small steps so as to explore in details the state diagram. Because of the dispersion in cell properties and of the few numbers of cells seen in each picture, the time evolution of characteristic properties such as the mean apparent aspect ratio is rather noisy and small transition times could not be measured. Therefore, we ran a few experiments with more abrupt increase or decrease of $\dot{\gamma}$ .

Some transitions in the high $\eta_0$ case are shown in Fig. \ref{fig:trans2+7}, by considering the mean apparent aspect ratio $\langle r_a\rangle(t)$, where the average is made over all cells on each image.

\begin{figure}
\begin{center}
  \includegraphics[width=\columnwidth]{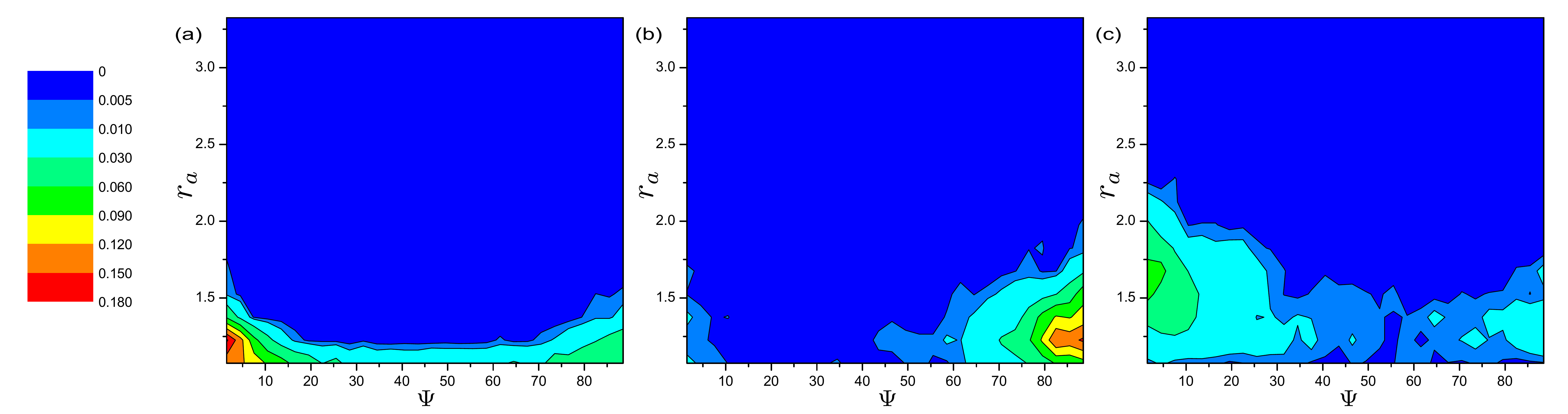}
\caption{Maps of probability densities in the $(\Psi,r_a)$ space along a transition towards the hysteresis zone in the decreasing shear rate case. $\eta_0=25$ mPa.s. Left panel corresponds to the 47 s - 62 s interval in Fig. \ref{fig:trans2+7}(b) ($\tau=0.73$ Pa). Middle panel  corresponds to the 62 s - 74 s interval right after the transition to  $\tau=0.073$ Pa. Right panel to  the 195 s - 207 s interval, still at  $\tau=0.073$ Pa. Left and right panel correspond to the initial and final states between which the arrow on top of Fig. \ref{fig:data2+7} indicates a transition. Middle panel is the transient state.}\label{fig:intermittent}
\end{center}
\end{figure}

Almost all situations, even when starting from rest (Fig. \ref{fig:trans2+7}(a)), lead to transition times that span between 10 and 70 ms. This is in agreement with stop-and-go experiments in a channel performed in  \cite{prado15}. The only longer transition time found here is when decreasing $\dot{\gamma}$ so as to switch from tank-treading regime to coexistence between tank-treading and flipping: see final transition in Fig. \ref{fig:trans2+7}(b). This transition is indicated by an arrow on top of Fig. \ref{fig:data2+7}. Transitions within the tank-treading regime (the two first transitions in  Fig. \ref{fig:trans2+7}(b)) do not exhibit such a long transition time: we conclude that this long time is not associated with the relaxation of the cells that remained in the tank-treading regime but rather to the regime transition. At first sight, this may be related to the intermittent regime observed in \cite{dupire12}, where successions of high angle orbits and tank-treading motions are observed before  stabilization towards a low angle flipping regime. Indeed, right after the change of shear rate, the probability distribution in the $(\Psi,r_a)$ space is quite different from the final distribution, which is characterized by orbits between 0 and 40$^\circ$ and a population of cells in tank-treading regime with apparent angle around 90$^\circ$ while in the initial state, the tank-treading cells have an angle either around 0$^\circ$ or around 90$^\circ$ (see Fig. \ref{fig:intermittent}). In this transient state, (Fig. \ref{fig:intermittent}, middle panel), angles between 50$^\circ$ and $90^\circ$ are dominant with aspect ratio that are higher than that of the tank-treading cells. This is a clear signature of the presence of the flipping orbits with high angles, that eventually disappear within a time of order one minute.

In \cite{dupire12}, it was noted that the cells do not get much deformed during transitions, while in \cite{levant16}, important deformations are observed during the intermittent regime. On the other hand, in this latter study, no off-plane motion is observed. This may suggest that off-plane motion is an alternative scenario to in-plane motion with high deformations, and that the different flow geometry in \cite{levant16} favors in-plane motions. From our data, it is difficult to validate any scenario, because of the strong dispersion of behaviors within a short time lapse.

Note finally that even the presence of these transient orbits become statistically negligible for smaller steps of shear rates. In that case, the number of cells involved in the transition would be so small that their contribution in the distribution could not be detected accurately. This limit is the consequence of a statistical study led on a population whose mechanical properties span on a wide range.

Intermittent behavior has also been reported in \cite{abkarian07} in the increasing shear rate case (but no rolling as observed), as in some numerical studies  \cite[see][]{cordasco14,peng14}. Here, we do not observe such a regime: transitions are sharp, even when directed towards the heart of the hysteresis zone (see, e.g. the transition from 0 to 0.37 Pa in Fig. \ref{fig:trans2+7}(a)): the suspension converges quasi-immediately towards its final state, where no signature of tumbling is reported.

In \cite{bitbol86}, a transition time towards rolling of order $100\times \dot{\gamma}^{-1}$ is measured. Here, the transition from 0 flow to $\tau=0.37$ Pa corresponding to a shear rate of 15 s$^{-1}$ is of 100 ms, much lower than $100/15\sim
 7 $s. \\

\begin{figure}
\begin{center}
  \includegraphics[width=\columnwidth]{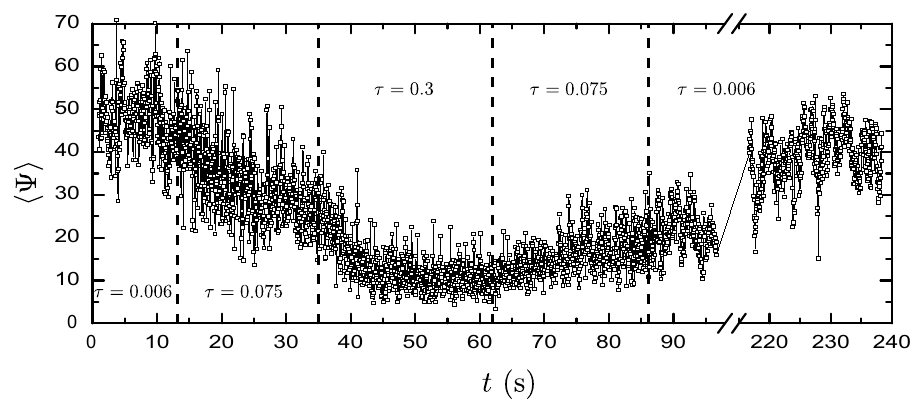}
\caption{Examples of transition dynamics for parameter $\langle \Psi \rangle$ after shear stress sharp increases  and decreases. $\eta_0=1.5$ mPa.s. Shear stresses $\tau$ are in Pa. Vertical dotted lines correspond to the shear stress instantaneous switches (by changing the shear rate).}\label{fig:trans0}
\end{center}
\end{figure}

\begin{figure}
\begin{center}
  \includegraphics[width=\columnwidth]{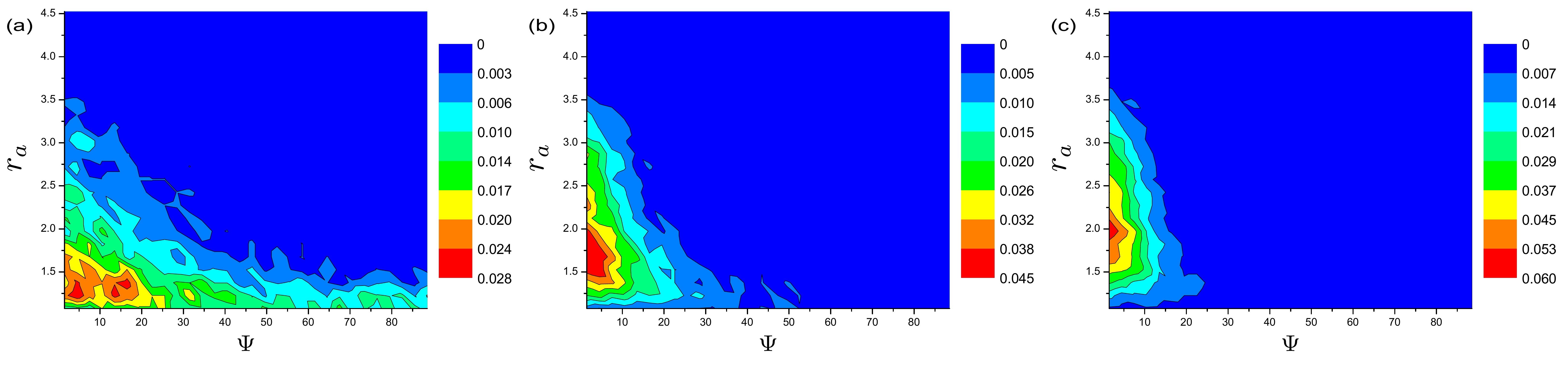}
\caption{Maps of probability densities in the $(\Psi,r_a)$ space along a transition towards rolling  in the increasing shear rate case, $\eta_0=1.5$ mPa.s. Left panel corresponds to the 25 s - 35 s interval in Fig. \ref{fig:trans0} ($\tau=0.075$ Pa). Middle panel  corresponds to the 35 s - 50 s interval right after the transition to  $\tau=0.3$ Pa. Right panel to the 50 s - 60 s interval, still at  $\tau=0.3$ Pa. Left and right panel correspond to the initial and final states between which the arrow on top of Fig. \ref{fig:data0}(c) indicates a transition}\label{fig:transit0detail}
\end{center}
\end{figure}

Some transitions for low $\eta_0$ are shown in Fig. \ref{fig:trans0}. Here, we consider the apparent angle as an indicator, as it varies more strongly than the apparent aspect ratio. Transition times are of order 10 s and more, and this result is not in  agreement with that of Bitbol either: the transition between $\tau=0.075$ Pa and $\tau=0.3$ Pa, that is detailed in the density maps of Fig. \ref{fig:transit0detail} corresponds to a mean angle continuous switching from 25$^\circ$ to around 10$^\circ$ and corresponds clearly to a transition towards rolling, as seen also in Fig. \ref{fig:data0}(c). The corresponding shear rates are 50 and 200 s$^{-1}$ so one should expect from  \cite{bitbol86} (and also from \cite{cordasco14_2}) a transition time of order 1 s, one order of magnitude lower than what we see here.

Finally, from a methodology viewpoint, the observed transition time of sometimes one minute for very $\dot{\gamma}$ decrease validates the method for exploring steady states: shear rates were varied by small steps, many movies separated by 15 s were taken and the first one was taken around 30 s after the change of shear rate.

\section{Conclusion}

We have proposed a complete description of the dynamics of a large, dilute population of red blood cells, that takes into account the dispersion in size and mechanical properties within this population. This dispersion, which is partly documented in the literature results in large ranges for the transition thresholds as commented here.

Hysteretic behavior for the transition between rolling and tank-treading has been highlighted, in the case of  suspending fluid of viscosity 25 mPa.s. We confirm the presence of intermittent regimes during the transition from tank-treading to the rolling regime, though quantifying this dynamics  would require another more complete study. In contrast with \cite{dupire12}, cells leaving the tank-treading regime seem to switch to small angle orbits, and not to any stable orbit. The characteristics of the stable orbits are the same whatever the direction of variation of the shear rate. Orbits with angles larger than a threshold of order 50$^\circ$ are  unstable. The transition to these orbits does not occur  at the same stress than the transition from these orbits, a feature which has not been observed before.

In the low and physiological viscosity case, this hysteresis between stable and unstable orbits was clearly identified. The characteristics of the stable orbits, as well as the range of existence of only stable orbits, are quite similar in the low and high viscosity cases.  While in the high viscosity case rolling cells eventually switch to tank-treading motion, some of them become stomatocytes in the low viscosity case. These two transitions are observed for stresses  that are not that different (of order 0.05 to 0.1 Pa). \\

One of the consequence of the deformability of cells under flow is the creation of  fore-aft asymmetry that allows lateral migration in the vicinity of walls, even in Stokes flow \cite[see][]{Olla97,coupier08}. In \cite{grandchamp13}, the migration of red blood cells in the vicinity of the walls of the same chamber as the one used here was studied. It was found that, within the studied range of viscosity and shear rates, there is no other effect of the shear rate than that of scaling the time. In particular, for $\eta_0=6.1 $ mPa.s, the curves $y'=f(\dot{\gamma} t)$ collapse for $\dot{\gamma}$ between 10 and 50 s$^{-1}$ that is, $\tau$ between $0.06$ and 0.3 Pa. On the other hand, in that stress range, we can expect from our results to observe a change in accessible orbits. We conclude that this change in behavior does not strongly affect the surrounding flow  and does not modify the resulting lift force. This does not mean that stresses do not influence the lift: at fixed shear rate and upon an increase of suspending fluid viscosity,  the lift force increases \cite[see][]{grandchamp13}. More studies on wider ranges of shear rate  are necessary to explore this issue. 

From a rheology viewpoint, it has been shown in \cite{vitkova08} that the contribution of cells to fluid viscosity depends on their regime in a non-trivial way. The existence of hysteresis loops in the dynamics indicates that blood viscosity may depend on the flow history, a fact that has never been reported (nor sought) to our knowledge.\\

While more and more numerical simulations approaches are found in the literature, with models mostly validated through comparison with simple configurations such as stretching experiments, we hope that this study based on statistically significant data will provide a reliable benchmark that will help explore issues hardly  solvable through sole experiments, such as that of the nature and properties of cytoskeleton, the impact of cell properties on rheology, and on the structuring of the suspension. It may also support the development of adequate viscoelastic models such as in \cite{mendez18} that would account for the important features we highlighted here, such as the presence of two hysteretic transitions and the robustness of the dynamical regimes against external viscosity variation, as long as stress is conserved.\\

 This experimental work benefited from instrumentation and a collaboration developed in the framework of microgravity experiments supported by CNES (parabolic flights) and ESA (sounding rockets). The authors would like to thank CNES and ESA  for providing these opportunities and for supporting research, and B. Polack from CHU Grenoble Alpes for help and discussions.

\bibliographystyle{jfm}
\bibliography{bibliovesicules}

\end{document}